\documentclass[%
 reprint,
prb,
]{revtex4-2}

\usepackage{dcolumn}
\usepackage{amsmath}
\usepackage{float}
\usepackage{graphicx}
\usepackage{siunitx}
\usepackage{bm}

\renewcommand{\eqref}[1]{Eq.~(\ref{eq:#1})}
\newcommand{\Eqref}[1]{Equation~(\ref{eq:#1})}
\newcommand{\citeasnoun}[1]{Ref.~\cite{#1}}
\newcommand{\figref}[1]{Fig.~\ref{fig:#1}}
\newcommand{\Figref}[1]{Figure~\ref{fig:#1}}
\newcommand*{\vect}[1]{\bm{#1}}
\newcommand*{\cc}[1]{{#1}^*}
\newcommand*{\Ev}{\vect{E}}
\newcommand*{\xv}{\vect{x}}
\newcommand*{\uv}{\vect{u}}

\newcommand*{\Ens}{\vect{E}_{i}}

\newcommand*{\epsv}{\hat{\vect{\varepsilon}}}
\newcommand*{\epsns}{\epsv_{i}}
\newcommand*{\kv}{\vect{k}}

\newcommand*{\cns}{c_{i}}
\newcommand*{\cnst}{c^\textrm{tar}_{i}}
\newcommand{\SM}{SM}
\renewcommand{\Re}{\operatorname{Re}}

\begin{document}

\title{High-NA Achromatic Metalenses by Inverse Design}

\author{Haejun Chung}
\author{Owen D. Miller}%
\affiliation{Department of Applied Physics and Energy Sciences Institute, Yale University, New Haven, Connecticut 06511, USA}%





\begin{abstract}
    We use inverse design to discover metalens structures that exhibit broadband, achromatic focusing across low, moderate, and high numerical apertures. We show that standard unit-cell approaches cannot achieve high-efficiency high-NA focusing, even at a single frequency, due to the incompleteness of the unit-cell basis, and we provide computational upper bounds on their maximum efficiencies. At low NA, our devices exhibit the highest theoretical efficiencies to date. At high NA---of 0.9 with translation-invariant films and of 0.99 with ``freeform'' structures---our designs are the first to exhibit achromatic high-NA focusing.
\end{abstract}


\maketitle

\section{Introduction}
Metasurfaces~\cite{yu2014flat,aieta2015multiwavelength} are patterned optical thin films that offer the possibility of manipulating light, for applications from holography to lenses, with precision equal to or greater than their bulky conventional-optics counterparts. For metasurface lenses, i.e., metalenses, a now-standard approach of ``stitching'' together wavelength-scale resonators into a larger device has demonstrated the possibility of focusing~\cite{aieta2015multiwavelength,khorasaninejad2015achromatic,avayu2017composite,shrestha2018broadband,chen2017gan,paniagua2018metalens,chen2018broadband,wang2018broadband}, but has suffered from narrow-bandwidth operation, low-numerical-aperture restrictions, and/or low focusing efficiencies. In this paper, we prove that unit-cell designs \emph{cannot} have high efficiency for high numerical apertures; conversely, we theoretically demonstrate that ``inverse design,'' a large-scale computational design technique optimizing all geometrical degrees of freedom~\cite{bendsoe1999material,bendsoe2001topology,neves2002topology,molesky2018inverse,miller2012photonic,pestourie2018inverse,camayd2018scaling,piggott2015inverse,jensen2011topology,yang2009design,lin2018topology} can discover high-numerical-aperture (``fast'') lenses that operate over visible bandwidths with best-in-class focusing efficiencies. Inverse design enables rapid computation of gradients with respect to arbitrarily many geometrical degrees of freedom; our implementation using a minimax formulation of the design criteria identifies fabrication-ready designs that can simultaneously achieve the large bending angles of high-NA lenses with the broad-bandwidth control that is necessary. Our results demonstrate the capabilities of adjoint-based approaches for superior design, and the emerging possibilities for combining multiple high-efficiency, hard-to-achieve functionalities in a single metasurface.

\begin{figure*}[t]
\centering
\includegraphics[width=1.0\linewidth]{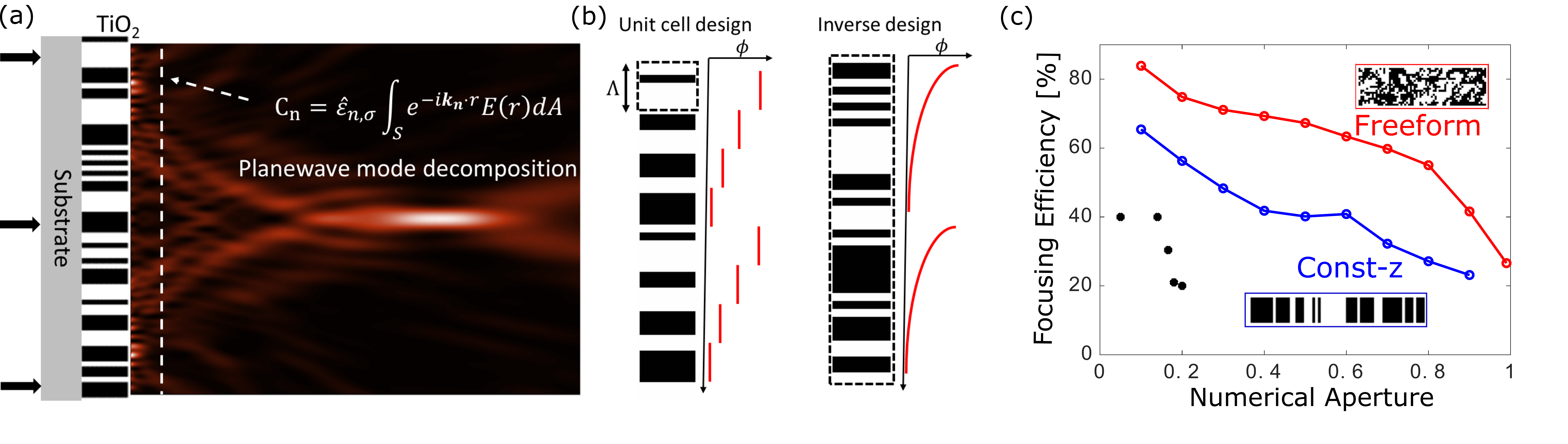}
\caption{Metalens inverse design. (a) The figure of merit for optical focusing is computed at an exit plane of the metalens (dashed line). (b) Schematic comparison of the unit-cell design approach, whereby a single (ideal) phase is fixed, using an incomplete basis periodic-boundary diffraction orders, versus large-area inverse design, whereby the full device scattering properties are incorporated into the design. (c) Compilation of broad-bandwidth inverse-designed metalenses using ``freeform'' (red) and constant-$z$ (``Const-z'') geometries, showing relatively high efficiencies for numerical apertures ranging from 0.1 to 0.99. The black circles indicate recently published results~\cite{wang2018broadband,chen2018broadband,mohammad2018broadband} with similar bandwidths, which operate only in the low-NA regime.}
\label{fig:schematic}
\end{figure*}

Metasurface designs and implementations have demonstrated the core functionality required for many applications: holography~\cite{zheng2015metasurface,ni2013metasurface,zhao2016full}, retroreflection~\cite{arbabi2017planar}, flat lenses~\cite{chen2018broadband,paniagua2018metalens,chen2017gan,shrestha2018broadband,khorasaninejad2016metalenses,wang2018broadband} and tunable optical components~\cite{kamali2016highly}. Yet that functionality is typically highly restricted in bandwidth, angular acceptance, and/or numerical aperture. Underlying these restrictions is the intuitive ``unit cell'' approach that pioneered initial metasurfaces designs~\cite{yu2014flat,aieta2015multiwavelength,khorasaninejad2016polarization}, whereby large-area films (tens to thousands of wavelengths in diameter) are constructed from libraries of wavelength-scale unit cells whose outgoing-wave phases, under ``locally periodic'' boundary conditions, are optimized for a single or few frequencies. This approach has been a crucial first step for designing large-area structures with complex patterning. Yet in the case of metalenses, which require precise control of potentially rapidly varying wavefronts, no metalens designs to date have achieved broad bandwidth and high numerical aperture. And as we show here, the assumption of local periodicity is incompatible with the requirements of high-NA focusing, wherein the amplitude and phase of the outgoing field must undergo rapid variations. Using a basis-projection approach we show that an assumption of local periodicity necessarily entails large focusing-efficiency losses, even at a single frequency.

To circumvent the limitations of the resonator-based unit-cell approach, we instead use inverse design to discover full-wave structures that simultaneously achieve broad bandwidth, high numerical aperture, and high efficiency. Inverse design centers around computing gradients with respect to large numbers of structural degrees of freedom by ``adjoint''-based methods. Adjoint-based sensitivities trace their roots to control theory~\cite{Pontryagin1962,cea1973quelques,Pironneau2012,Giles2000}, and have since been used for rapid, efficient optimization in circuit theory~\cite{Director1969}, aerodynamics~\cite{Jameson1988}, mechanics and elasticity~\cite{Bendsoe2013}, quantum dynamics~\cite{Demiralp1993,Rabitz2004}, and deep learning~\cite{Werbos1994,Rumelhart1986,LeCun1989}, where it is known as ``backpropagation.'' More recently it has emerged as a promising design tool for nanophotonics~\cite{yang2009design,jensen2011topology,frandsen2014topology} for applications including waveguide demultiplexers~\cite{lalau2013adjoint,piggott2015inverse,su2017inverse}, beam deflectors~\cite{sell2017large,callewaert2018inverse}, photonic bandgaps~\cite{men2014robust}, solar cells~\cite{ganapati2014light}, and many others. Preliminary studies have applied inverse design to metasurfaces~\cite{lin2018topology,callewaert2018inverse,shen2014ultra,molesky2018inverse}, including large-area metasurfaces~\cite{pestourie2018inverse,lin2019topology}, albeit thus far limited to isolated frequencies or metrics other than lens focusing (beam deflection, polarizers, etc.). 

In this work, we use inverse design to demonstrate broadband achromatic metalenses operating with relatively high efficiencies across the visible for both small and large numerical apertures as shown in \figref{schematic}. To fully explore the design space and the tradeoffs associated with numerical aperture, bandwidth, and efficiency, we primarily design two-dimensional dielectric profiles, while demonstrating that the methods scale to fully three-dimensional films. We make no periodic / unit-cell approximations, and indeed the designs that we discover often show rapidly varying spatial profiles; to ensure feasible computation times, we design devices in the 10--60$\lambda$ size range. With recent developments in fast electromagnetic solvers~\cite{Liu2016,Bruno2019}, one can anticipate applying this approach to devices orders of magnitude larger, with small and controllable errors, in the near future. The full-device optimizations enable us to overcome the efficiency losses and single-functionality limitations associated with assuming small, periodic unit cells, which are especially prominent at high NA. Our metalens designs demonstrate the capability for inverse design to lead the discovery of non-intuitive yet superior metalens devices.

\section{Unit-cell-Metalens Efficiency Limits}
\label{physics}
The prototypical approach to metalens design is the unit-cell method~\cite{khorasaninejad2015achromatic,khorasaninejad2016metalenses, wang2018broadband}, wherein the fields at the exit plane of the metalens are designed by breaking them into wavelength-scale ``unit cells'' that vary slowly and can be treated locally as periodic elements. This reduces the large-scale design problem into a large number of much smaller design problems, while simplifying the physics of the problem. In the prototypical case with a unit-cell period smaller than the wavelength, there is only a single outgoing diffracted-wave order for each cell, and the design problem simplifies to one of controlling the phase and amplitude of each cell's outgoing wave. Such a formulation extends to broadband metalens operation, where all of the complexity of large-area, many-frequency operation can be simplified to control of phase and its first two derivatives---group velocity and group velocity dispersion---at a single frequency~\cite{chen2018broadband}. Yet there are complex tradeoffs associated with the unit-cell approach. For any fixed fabrication tolerance, a larger period allows for more complex unit cells, but as the period increases beyond the wavelength, higher-order modes are introduced, which if not controlled can lead to significant efficiency losses. Conversely, smaller periods reduce the number of geometrical degrees of freedom, particularly if the local-periodicity assumption is to hold.

\begin{figure*}[t]
\centering
\includegraphics[width=\linewidth]{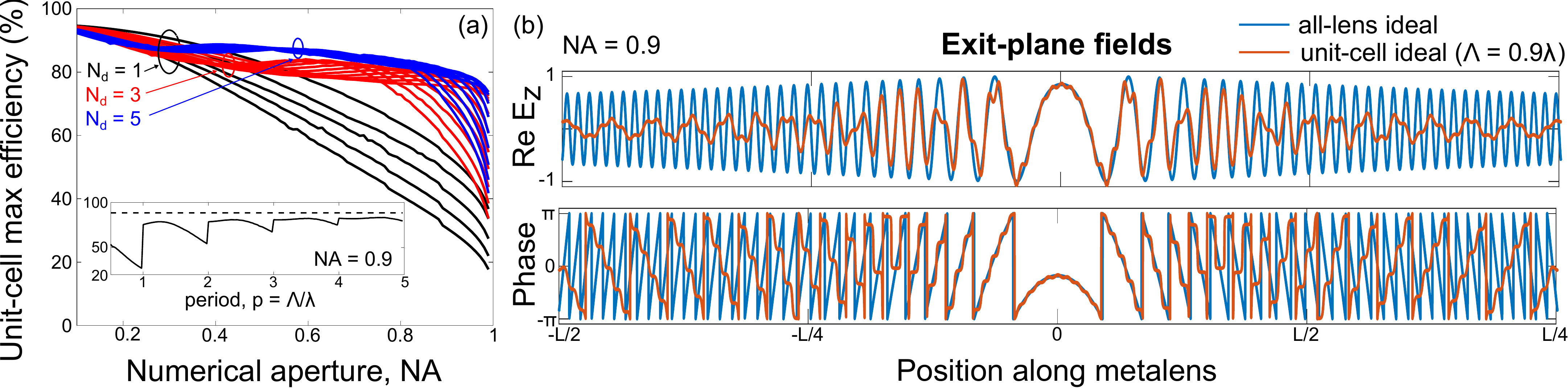}
\caption{(a) Upper bounds to the single-frequency focusing efficiency of a metalens designed by a unit-cell approach. (Shown for a 2D metalens with diameter $L = 100\lambda$.) For high numerical apertures, there is a significant efficiency loss incurred by the unit-cell approach, especially for unit-cell periods $\Lambda$ close to but smaller than the wavelength $\lambda$, where there is a single diffraction order ($N_d = 1$) and the best theoretical efficiency may only be 20\%. Inset: variation of the maximum efficiency as a function of period, with abrupt discontinuities as new orders appear. (b) Optimal fields, real part (upper) and phase (lower), of the ideal metalens (blue) and the ideal unit-cell design (orange), showing that non-periodic nature of the focusing fields leads to amplitude and phase inaccuracies in the unit-cell approach.}
\label{fig:phase_figure}
\end{figure*}

We use a modal-decomposition analysis to derive bounds on focusing efficiencies for metalenses designed by a unit-cell approach, at any given frequency. We show that even without incorporating deviations from the uncontrolled assumption of periodic boundary conditions, a high numerical aperture \emph{necessarily} incurs significant efficiency losses in the unit-cell approach.

Consider a lens focusing light to a focal point. At the exit plane of the lens, the optimal fields $\Ev^{\rm ideal}$ are the time-reversed conjugates of those radiated from a dipole at the focal point. (The dipole polarization may be selected for maximum efficiency.) In a unit-cell approach, designing a metalens by stitching together library elements simulated with periodic boundary conditions (over period $\Lambda$) is equivalent to designing the field $\Ev^{\rm UC,ideal}$ as a linear combination of basis functions $\uv_{m,\Lambda}(\xv)$ that are the diffraction orders of each ``periodic'' unit cell:
\begin{align}
    \Ev^{\rm UC,ideal}(\xv) = \sum_{m=1}^{N_d N_{uc}} c_{m,\Lambda} \uv_{m,\Lambda}(\xv),
    \label{eq:Euci}
\end{align}
where $N_{uc}$ is the number of unit cells, $N_d$ is the number of (polarization-resolved) diffraction orders per unit cell, and the $\Lambda$-periodic diffraction orders $\uv_{m,\Lambda}$ are given by
\begin{align}
    \uv_{m,\Lambda}(\xv) = \frac{1}{\sqrt{A_\Lambda}} \epsv_m e^{i\left[2\pi (a_m x + b_m y) / \Lambda + k_{zm}z\right]} H_m(\xv),
    \label{eq:bf}
\end{align}
where $\epsv_m$ is the unit-vector polarization, $A_\Lambda$ is the unit-cell area, $H_m(\xv)$ a Heaviside step function that is nonzero only for the unit cell corresponding to mode index $m$, the variables $a_m$ and $b_m$ are integers, and $k_{zm}$ is the appropriate forward-propagation wavevector component such that the total wavevector magnitude is $\omega/c$, for frequency $\omega$. The basis functions $\uv_{m,\Lambda}$ form an orthonormal set of functions over the area $A$ of the device, i.e. $\int_{A} \uv_{m,\Lambda}^\dagger \uv_{n,\Lambda} = \delta_{mn}$. 

Every unit-cell metasurface design~\cite{khorasaninejad2016polarization,khorasaninejad2016metalenses,zheng2015metasurface,shrestha2018broadband,chen2018broadband,wang2018broadband} is implicitly using \eqref{Euci} to describe the outgoing fields. The true ideal field is the time-reversed field from a dipole at the focal point, which we denote $\Ev^{\rm ideal}$. The best possible unit-cell design is the one that minimizes the difference between $\Ev^{\rm UC,ideal}$ and $\Ev^{\rm ideal}$; for example, the field that minimizes $\|\Ev^{\rm ideal} - \Ev^{\rm UC,ideal}\|_2^2$. (Often only the phase is optimized and the efficiency decreases further; since we are interested in upper bounds, we assume the ideal scenario.) By the orthogonality of the basis functions in \eqref{bf}, there is a set of unit-cell coefficients $c_{m,\Lambda}$ that minimize this least-squares quantity, given by (cf. {\SM})
\begin{align}
    c_{m,\Lambda} = \int \uv_{m,\Lambda}^\dagger(\xv) \Ev^{\rm ideal}(\xv).
    \label{eq:cmLam}
\end{align}
From an infinite library of unit cells, then, stitching together those with coefficients closest to the distribution of \eqref{cmLam} represents the ideal unit-cell design. Yet even if such library elements exist, the simulated performance of the full device falls short in efficiency, in part for a simple reason: generically, the field $\Ev^{\rm UC,ideal}$ of \eqref{Euci} is not a valid solution of Maxwell's equations.

We can semi-analytically determine the closest approximation to \eqref{Euci} which \emph{is} a valid solution of Maxwell's equations. At the exit plane of the metalens, we can write the fields not as a superposition of unit-cell basis functions, but instead as a linear combination of plane waves:
\begin{align}
    \Ev^{\rm UC}(\xv) = \sum_{m=1}^{N_{pw}} c_{m,L} \uv_{m,L}(\xv),
    \label{eq:Euc}
\end{align}
where the $\uv_{m,L}$ have the same form as in \eqref{bf}, but replacing the period $\Lambda$ with the device diameter $L$, and removing the Heaviside function. (Since we are interested in large metalenses with $L \gg \lambda$, we simplify the discretization to $N_{pw}$ waves in \eqref{Euc} by assuming periodic boundary conditions, which have no effect in the large-$L$ limit.) The Maxwell field $\Ev^{\rm UC}$ that is closest to $\Ev^{\rm UC,ideal}$ can be found by minimizing the squared two-norm $\|\Ev^{\rm UC,ideal} - \Ev^{\rm UC}\|_2^2$; by the orthogonality of the $\uv_{m,L}$ (analogous to the argument above), the coefficients $c_{m,L}$ will be given by
\begin{align}
    c_{m,L} &= \int \uv_{m,L}^\dagger(\xv) \Ev^{\rm UC,ideal}(\xv) \nonumber \\
            &= \sum_{n=1}^{N_d N_{uc}} \int \uv_{m,L}^\dagger(\xv) \uv_{n,\Lambda}(\xv) \int \uv_{n,\Lambda}^\dagger(\xv') \Ev^{\rm ideal}(\xv').
            \label{eq:cmL}
\end{align}
Rearranging \eqref{cmL} would yield an intuitive term sandwiched in the middle: $\sum_n \uv_{n,\Lambda}(\xv) \uv_{n,\Lambda}^\dagger(\xv')$, which is the orthogonal projector~\cite{Trefethen1997} onto the unit-cell basis functions. If the unit-cell basis were complete, i.e. it spanned the space of all possible solutions, then that term would be a delta function, \eqref{cmL} would simplify to the overlap of the ideal fields with the plane-wave basis functions, and 100\% efficiency would be possible. However, the unit-cell basis is not complete: the periodic orders of the unit cell do not span the space of all possible solutions, and this shortcoming necessarily leads to efficiency losses.

\Eqref{cmL} can be evaluated for any unit-cell period $\Lambda$ and numerical aperture, with the resulting focusing-plane fields computed by \eqref{Euc}. The intensity of the fields at the focal point, relative to those of the ideal field $\Ev^{\rm ideal}$, thereby represents an upper bound on the maximum designable efficiency by a unit-cell approach. We plot the upper bound in \figref{phase_figure}(a), showing the steep dropoffs in maximum efficiency as numerical aperture increases. \Figref{phase_figure}(b) shows the ideal and unit-cell-ideal fields at the exit plane of the metasurface, with the latter unable to capture the necessarily rapid variations in phase and amplitude. We emphasize that the efficiency upper bound accounts only for the incompleteness of the unit-cell basis (including effects such as insufficient spatial sampling of the phase~\cite{swanson1989binary,kamali2018review}); it assumes no further losses due to multiple-scattering effects between non-identical neighboring cells (that violate the local-periodicity assumption), and thus almost certainly \emph{over}estimates the maximum efficiency possible in a unit-cell approach. Even so, these modal-decomposition bounds predict significant efficiency losses for unit-cell-based high-numerical-aperture metalenses. An alternative is to design the entire device at once, a task of significant complexity where inverse design may be ideal.

\section{Metalens inverse design framework}\label{Principle}
Inverse design requires specification of a figure of merit as well as the geometrical degrees of freedom. To determine the focusing properties of a given metalens geometry, we compute the fields at an exit plane of the metasurface and project them to the far field. The plane-wave decomposition at the exit plane is given by 
\begin{equation}
    \Ev(\xv) = \frac{1}{A} \sum_{i}\cns\epsns e^{i\kv_i \cdot \xv} = \sum_i c_i \Ev_i(\xv)
    \label{eq:Esum}
\end{equation}
where $\kv_i$ the wavevector of each plane wave, $i$ a discretized order comprising angle and polarization, and $\epsns$ is the polarization unit-vector. For the figure of merit (FOM), we use a measure of the overlap between the exit-plane electric field, with field coefficients $c_i$, with a ``target'' field, the time-reversed (conjugated) field emanating from an electric dipole at the desired focal point, with field coefficients $\cnst$:
\begin{equation}
    \mathcal{F} = \frac{1}{2}\sum_{i} |c_i^* \cnst|^2,
    \label{eq:fom}
\end{equation}
where the asterisk denotes complex conjugation. \Eqref{fom} is a function of frequency, and independent of a global phase in the vector of coefficients $\cnst$, since we do not want to fix the phase of the field at the focal point. To optimize broadband focusing, one could maximize the frequency average of the overlap in \eqref{fom}, but optimizations of the average tend to converge to solutions with very good performance at isolated frequencies but poor performance across much of the bandwidth of interest. Instead, a more robust approach is to optimize the worst-case performance across the bandwidth of interest, ensuring that every frequency achieves at least a modest if not exceptional level of efficiency. Such approaches generally fall under the umbrella of ``minimax'' optimizations~\cite{ganapati2014light}, though since our natural metric is one of maximization, we technically use \emph{maximin} over the geometrical degrees of freedom and the frequency range of interest:
\begin{equation}
    \max\limits_{\rm geo}[\min\limits_{\omega}\mathcal{F}(\Ev)].
    \label{eq:maximin}
\end{equation}
In all of the designs presented below, we use TiO$_2$ as the metalens material, incorporating its dispersion across visible wavelengths~\cite{palik1998handbook}. For the geometrical degrees of freedom, we allow the density of TiO$_2$ to vary between 0 and 1 at every point in the structure (a ``topology optimization'' approach), and then add penalty functions~\cite{neves2002topology} to \eqref{maximin} to enforce a binary-material constraint (ultimately converging to densities of 0 and 1 at every point). In some cases we also enforce the constraint of constant permittivity in one ($z$) direction, akin to a slab that can be fabricated by conventional lithography techniques. 

The critical step in inverse design is the efficient computation of the gradient with respect to the arbitrarily many geometrical degrees of freedom, to discover efficient updates from one geometry to the next. Instead of \emph{independently} simulating every geometrical perturbation, by reciprocity (or its generalization for nonreciprocal media~\cite{kong1975theory}), one can use a single coherent set of dipole sources, with spatially dependent amplitudes determined by the form of the figure of merit, to compute an ``adjoint'' field that in one simulation provides information about all possible perturbations~\cite{johnson2002perturbation}. For a metric with the form of \eqref{fom}, the current sources for the adjoint simulation are given by $\vect{J}_{\rm adj} = -i\omega \vect{P}_{\rm adj} = -i\omega \partial \mathcal{F} / \partial \Ev$ (\citeasnoun{miller2012photonic}), which on the exit plane of the metalens are given by:
\begin{align}
    \vect{J}_{\rm adj}(\xv) &= -i\omega \frac{\partial \mathcal{F}}{\partial \Ev} \nonumber \\
                            &= -\frac{i\omega}{2} \sum_{i} \cns |\cnst|^2 \Ens^{*}(\xv).
                            \label{eq:Jadj}
\end{align}
The derivative of $\mathcal{F}$ with respect to $\Ev$ is computed by inverting \eqref{Esum}, i.e. $c_i = \int_A \cc{\Ev}_i(\xv) \cdot \Ev(\xv)$, where $A$ is the area of the exit plane. Per \eqref{Jadj}, the adjoint sources can be physically interpreted as a weighted combination of time-reversed plane wave modes ``back-propagated'' (as in deep-learning neural networks~\cite{Rumelhart1986,LeCun1989,Werbos1994}) into the design region of the metalens. Once the adjoint fields are known, the gradient of $\mathcal{F}$ with respect to permittivity perturbations $\delta \varepsilon$ can be computed from the variation $\partial \mathcal{F} / \partial \varepsilon(\xv) = \Re \left[ \Ev(\xv) \cdot \Ev_{\rm adj}(\xv) \right]$. These derivatives can then be used in gradient descent or other common optimization algorithms to discover an optimal design.

The plane-wave-decomposition approach to metalens inverse design is depicted in \figref{schematic}(a). We simulate every geometry with \textsc{Meep}, a free-software implementation~\cite{oskooi2010meep} of the finite-difference time-domain (FDTD) method~\cite{taflove2013advances}. We consider broad-bandwidth response for 450--700nm wavelengths, representing a large 43\% relative bandwidth. We optimize the fields at 20 wavelengths in this range and verify that the optimal devices have a smooth spectral response at substantially higher resolution (120 wavelengths). (We found that optimizing with 10 frequency points or fewer may lead to highly oscillatory frequency response with wide deviations within the visible-frequency bandwidth at intermediate, non-optimized frequencies, as shown in the SM.) The maximin metric of \eqref{maximin} is not differentiable everywhere, caused by by frequency crossings whereby infinitesimal geometric perturbations cause the lowest-efficiency frequency to change. This can be handled by a standard optimization transformation to epigraph form~\cite{boyd2004convex}, in which a dummy scalar variable is maximized, and the value of $\mathcal{F}$ in \eqref{fom} at each frequency becomes an inequality constraint, but for metalenses that are many wavelengths in size, with many frequencies to be computed, this can require substantial data storage. Instead, for every geometry we select only the minimum-efficiency frequency's gradient and use that as the gradient for the more general FOM, which is the correct gradient everywhere that the FOM is differentiable. In theory such an approach could lead to oscillations and slow convergence, but in practice we see robust convergence to high-quality optima. Each iteration takes approximately 45 seconds on 25 cores in our computational cluster (Intel Xeon E5-2660 v4 3.2 GHz processors). The 
least-squares figure of merit rapidly improves then converges in about 200 iterations, after which the geometric penalization transforms the grayscale design to a binary one, which is a slow process (on the order of 2000 iterations) but which results in little-to-no efficiency losses.

\section{Achromatic metalens design}\label{Results}
In this section, we design metalenses across a range of low to high numerical apertures, achieving relatively high efficiencies for each optimal device. We assume two-dimensional devices with thicknesses of 250 nm and widths of 12.5 $\mu$m, with the latter chosen to be large enough ($\approx 20\lambda$) to require large phase variations but small enough for relatively fast optimizations (under 3 hours on the computers described above). In the {\SM} we demonstrate 2D designs with similar efficiencies for widths up to 60$\lambda$, as well as fully 3D metalens designs. For simplicity we do not incorporate a particular substrate, though again in the {\SM} we demonstrate similar performance for an optimized device on a substrate. We use TiO$_2$ for the material, and incorporate its material dispersion~\cite{palik1998handbook} by fitting a Lorentz--Drude model to the susceptibility. A transverse magnetic wave is used for the incident field. For the geometric ``pixels,'' we enforce a 25 nm minimum size to avoid non-robust, highly sensitive designs that are difficult to fabricate. We also verify robustness after optimization by simulating material imperfections in the optimized designs, which retain high efficiency even for moderate geometrical imperfections.

We have designed optimal metalensas at each of the 10 numerical apertures $0.1, \ldots, 0.9, 0.99$, with their efficiencies plotted in \figref{schematic}(d). We include further characteristic data for each in the SM; in the remainder of the section, we highlight and discuss three of the designs, at NA = 0.1, 0.9, and 0.99. All geometrical degrees of freedom for each design are included in the {\SM}.

\subsection{Low-NA, high-efficiency metalens}\label{High}

\begin{figure*}[!t]
\centering
\includegraphics[width=0.9\linewidth]{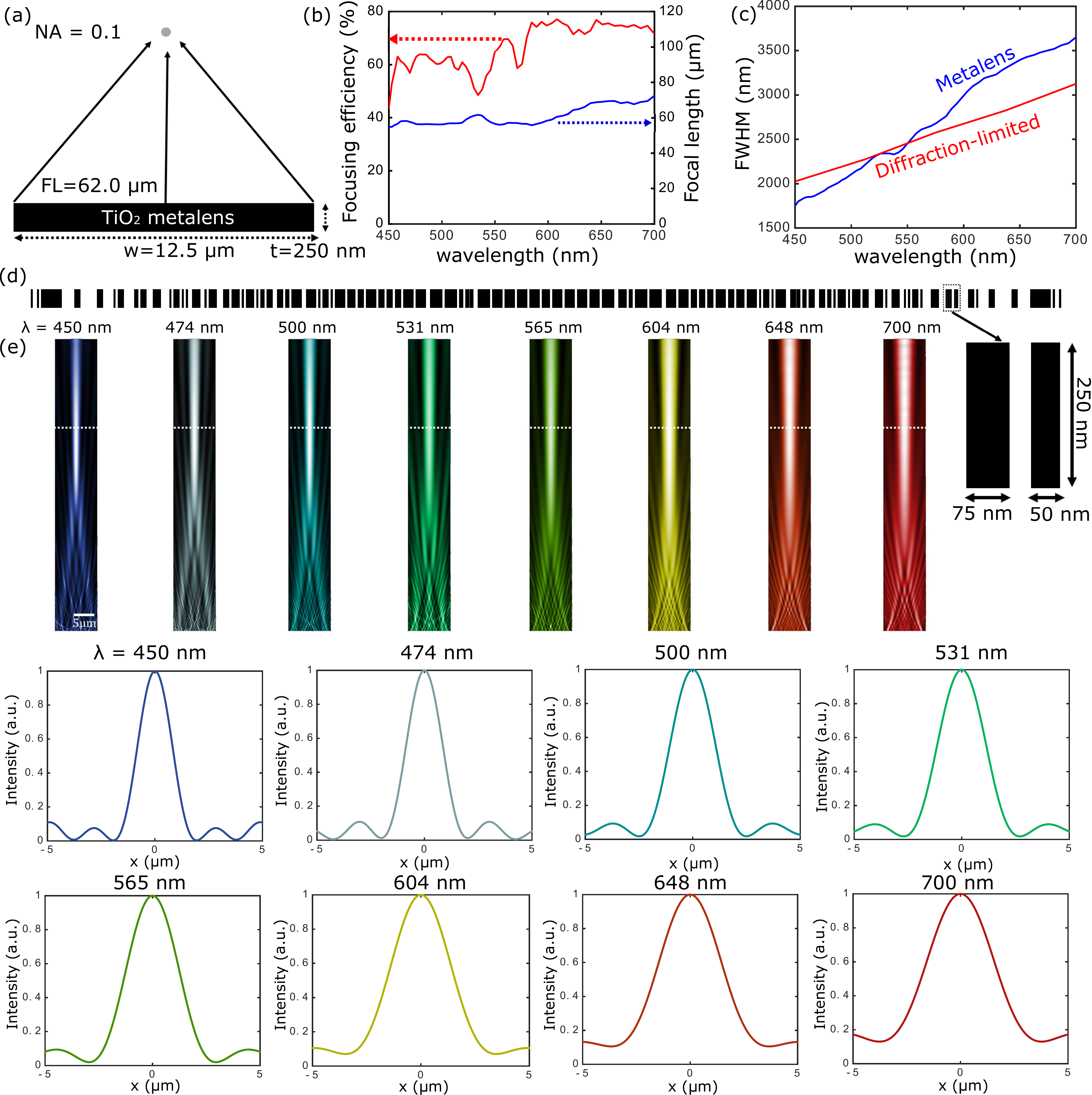}
\caption{NA = 0.1, high-efficiency achromatic metalens for visible wavelengths (450 -- 700 nm) using a lithography-compatible ``Const-z'' geometry. (a) Metalens dimensions. (Not to scale.) (b) Focusing efficiency and focal lengths of the optimized metalens. The average efficiency is 65\% over visible spectrum. The efficiency was calculated by power within the first minimum point divided by the incidence power. Calculated focal lengths remain within the depth of focus area at all wavelengths. (c) Full-width half-maximum (FWHM) of the optimized metalens and the corresponding Airy disk. (d) Optimized TiO$_2$-based metalens design. (e) Normalized intensity ($|E|^2$) profile. The curves are normalized intensity profiles at the focal plane.}
\label{NA01}
\end{figure*}

All broadband metalens designs to date~\cite{wang2018broadband,chen2018broadband,mohammad2018broadband,aieta2015multiwavelength,khorasaninejad2015achromatic,avayu2017composite,shrestha2018broadband} operate in the low-NA regime. Thus, we start with low-NA designs for a more direct comparison, and we demonstrate higher efficiencies than the current state-of-the-art.

For 12.5 $\mu$m width and 250 nm thickness, a 0.1-NA device has a 60 $\mu$m focal length, depicted in Fig.~\ref{NA01}(a). As shown in Fig.~\ref{NA01} (b), the designed device achieves an average focusing efficiency of 65\% and a maximum efficiency of 78\% over visible wavelengths (450 -- 700 nm). We use the same definition of focusing efficiency as \citeasnoun{chen2018broadband}: the ratio of the power concentrated within the region out to the first electric-field minimum divided by the total incident power (\emph{including} all reflection loss). The focal length remains nearly constant over wavelength. Any focal-length deviations remain within the depth of focus as measured by the full-width half-max of the field intensity. As shown in Fig.~\ref{NA01}(c), the focused electric fields are nearly diffraction-limited, and are even narrower than the diffraction limit at shorter wavelengths. (Imperfect focusing efficiencies enable sub-diffraction-limited fields~\cite{Ferreira2006,Shim2019}.) Figure.~\ref{NA01}(d) shows the optimized structure, comprising an alternating series of TiO$_2$ blocks and air holes. Considering the high-level geometric patterning, the density of TiO$_2$ increases from the lens edge to its center, as also seen in unit-cell-based designs~\cite{chen2018broadband, wang2018broadband}. On the other hand, very distinguishable from unit-cell designs, the small-scale geometrical variations are clearly non-periodic. Fig.~\ref{NA01}(e) shows the electric-field intensity profiles over many wavelengths, exhibiting clear high focusing efficiency; despite significant variations in the near-zone fields as the wavelength changes, each wavelength depicted (and all intermediate wavelengths not shown) produce focusing at the same focal plane. The images in Fig.~\ref{NA01} show the entire simulation region with no stray fields unaccounted for. Further increases in focusing efficiency would be possible with materials that have less dispersion than TiO$_2$. We enforced relatively large minimum features sizes (25 nm), for fabrication compatibility; smaller feature sizes may yield modest further efficiency improvements.

\subsection{High-NA metalens, NA = 0.9}\label{high NA}
\begin{figure*}[!t]
\centering
\includegraphics[width=0.9\linewidth]{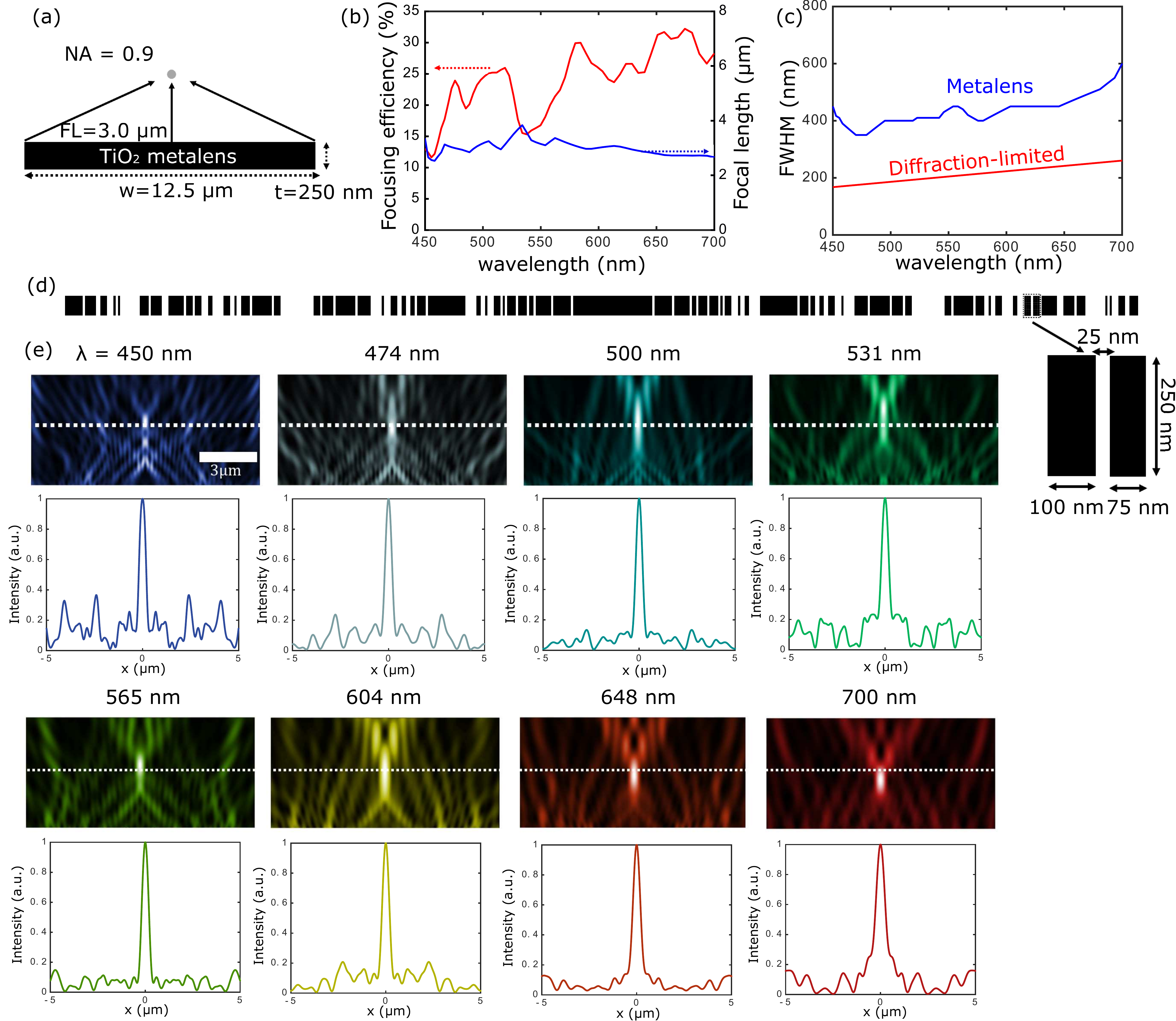}
\caption{High-NA (NA = 0.9) achromatic metalens for visible wavelengths. (a) Metalens dimensions. (b) Focusing efficiency and focal lengths of the optimized metalens. The average efficiency is 23\% over visible spectrum. (c) FWHM of the optimized metalens and the corresponding Airy disk. (d) Optimized metalens design. (e) Normalized intensity ($|E|^2$) profile everywhere (images) and at the focal plane (curves).}
\label{NA09}
\end{figure*}


In this subsection we design and demonstrate high-NA broadband achromatic metalenses. The optimal device, shown in Fig.~\ref{NA09}(a), has the same dimensions as the NA=0.1 lens while the target focal length is now 3 $\mu$m. The optimal-device focal length, shown in Fig.~\ref{NA09}(b), is nearly unchanged over visible wavelengths, and the focusing efficiency ranges from 13\% to 32\% with an average value of 23\%. As expected, the focusing efficiency is smaller than for the low-NA device, due to the increased difficulty of focusing light to a closer point over the same frequency bandwidth. The FWHM of the optimized metalens is slightly larger than that of the diffraction limited lens. As shown in Fig.~\ref{NA09}(d), the optimized device structure is again clearly non-periodic, with rapid changes in topology that accommodate the $\approx 12$ intervals of 0 to $2\pi$ phase change required at these dimensions. Normalized intensity ($|E|^2$) profiles are shown in Fig.~\ref{NA09}(e) for 8 selected wavelengths. The 450-nm-wavelength intensity profile (blue) shows a primary focal spot at the desired length of 3$\mu$m as well as a second focal spot at 1.9$\mu$m, which arises due to the relatively low focusing efficiency at that wavelength. To further improve efficiency of the high NA achromatic metalens, as we discussed in the previous section, one could use a much finer resolution of the structure up to fabrication limit. Multilayer metasurfaces~\cite{lin2018topology} are an alternative way of increasing the degree of freedom. This is illustrated by the fact that increasing the degrees of freedom with a ``freeform'' topology varying in the $z$ direction improves the average focusing efficiency from 23\% to 41\% for 0.9 NA metalenses, as shown in \figref{schematic}(d).

\subsection{Very-high-NA freeform metalens, NA=0.99}\label{freeform}
\begin{figure*}[!t]
\centering
\includegraphics[width=0.9\linewidth]{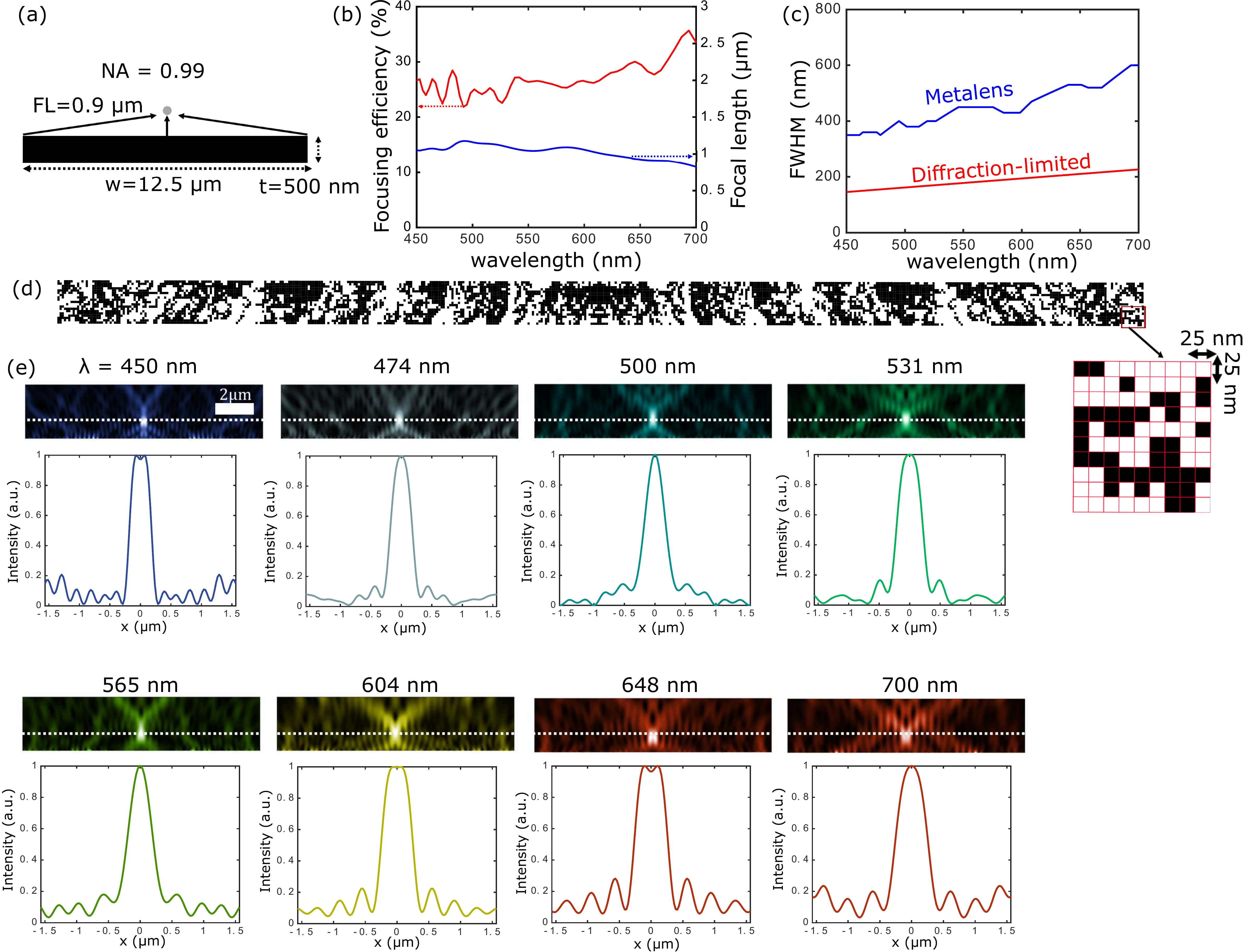}
\caption{Near-unity-NA (NA = 0.99) achromatic metalens achieved by inverse design of a freeform geometry. (a,b) Metalens dimensionsi, focusing efficiency, and focal lengths. (c) FWHM of the optimized metalens and the corresponding Airy disk. (d) Optimal design. (e) Normalized field-intensity profiles.}
\label{NA099}
\end{figure*}

In this section, we probe the extreme limit of high-NA design, relaxing the constant-$z$ topology constraint and designing a freeform-topology device that achieves NA = 0.99. The optimized device has 12.5 $\mu$m width, 500 nm thickness, and a focal length now of 0.9 $\mu$m, as depicted in Fig.~\ref{NA099}(a). The focal length shown in Fig.~\ref{NA09}(b) is again nearly constant over visible frequencies and the average focusing efficiency is 27\%. This result shows much higher efficiency compared to the same NA constant-$z$ structure metalens due to the increased degrees of freedom. The FWHM of the optimized metalens is slightly larger than that of the diffraction limited lens. As shown in Fig.~\ref{NA099}(d), the optimized structure exhibits diagonal patterns that presumably help redirect the light towards the nearby focal spot. Normalized intensity ($|$E$|$$^2$) profile is shown in Fig.~\ref{NA099}(e) for 8 selected wavelengths. It shows clear focal spots at the target focal plane (0.9 $\mu$m) over the visible spectrum. 

\section{Conclusions}
In this work, we have demonstrated the capability for inverse design to discover high-efficiency achromatic metalenses across the visible spectrum. We focused on 2D devices, to enable rapid systematic design of devices for ten different numerical apertures and two topologies (const-$z$ and freeform), as well as the many optimizations with varying initial conditions and hyperparameter choices for each case.  In the SM we demonstrate two generalizations: to larger diameters (up to 60$\lambda$) with very similar efficiencies, and to similarly sized 3D metalens devices. Our results corroborate those from the literature demonstrating that there is not a significant dropoff from 2D device designs~\cite{wang2016chromatic,aieta2015multiwavelength,khorasaninejad2015achromatic} to their 3D counterparts~\cite{wang2018broadband, chen2018broadband, mohammad2018broadband, kamali2016highly}.

Through inverse design, we have demonstrated the highest-efficiency low-NA achromatic metalenses to date, as well as the first theoretical demonstration of broadband high-NA structures. Breaking the local-periodicity assumptions of the unit-cell design approach results in significant enhancements in device efficiency and a new regime for performance. Scaling this approach to macroscopic length scales should be possible with emerging fast-solver techniques, ultimately leading to the possibility for numerous applications including wearable optical devices, microscopy and integrated optics. 

\section*{Funding Information}
H.~C. and O. D. M. were partially supported by the Air Force Office of Scientific Research under award number FA9550-17-1-0093.


\begin{thebibliography}{63}%
\makeatletter
\providecommand \@ifxundefined [1]{%
 \@ifx{#1\undefined}
}%
\providecommand \@ifnum [1]{%
 \ifnum #1\expandafter \@firstoftwo
 \else \expandafter \@secondoftwo
 \fi
}%
\providecommand \@ifx [1]{%
 \ifx #1\expandafter \@firstoftwo
 \else \expandafter \@secondoftwo
 \fi
}%
\providecommand \natexlab [1]{#1}%
\providecommand \enquote  [1]{``#1''}%
\providecommand \bibnamefont  [1]{#1}%
\providecommand \bibfnamefont [1]{#1}%
\providecommand \citenamefont [1]{#1}%
\providecommand \href@noop [0]{\@secondoftwo}%
\providecommand \href [0]{\begingroup \@sanitize@url \@href}%
\providecommand \@href[1]{\@@startlink{#1}\@@href}%
\providecommand \@@href[1]{\endgroup#1\@@endlink}%
\providecommand \@sanitize@url [0]{\catcode `\\12\catcode `\$12\catcode
  `\&12\catcode `\#12\catcode `\^12\catcode `\_12\catcode `\%12\relax}%
\providecommand \@@startlink[1]{}%
\providecommand \@@endlink[0]{}%
\providecommand \url  [0]{\begingroup\@sanitize@url \@url }%
\providecommand \@url [1]{\endgroup\@href {#1}{\urlprefix }}%
\providecommand \urlprefix  [0]{URL }%
\providecommand \Eprint [0]{\href }%
\providecommand \doibase [0]{https://doi.org/}%
\providecommand \selectlanguage [0]{\@gobble}%
\providecommand \bibinfo  [0]{\@secondoftwo}%
\providecommand \bibfield  [0]{\@secondoftwo}%
\providecommand \translation [1]{[#1]}%
\providecommand \BibitemOpen [0]{}%
\providecommand \bibitemStop [0]{}%
\providecommand \bibitemNoStop [0]{.\EOS\space}%
\providecommand \EOS [0]{\spacefactor3000\relax}%
\providecommand \BibitemShut  [1]{\csname bibitem#1\endcsname}%
\let\auto@bib@innerbib\@empty
\bibitem [{\citenamefont {Yu}\ and\ \citenamefont
  {Capasso}(2014)}]{yu2014flat}%
  \BibitemOpen
  \bibfield  {author} {\bibinfo {author} {\bibfnamefont {N.}~\bibnamefont
  {Yu}}\ and\ \bibinfo {author} {\bibfnamefont {F.}~\bibnamefont {Capasso}},\
  }\bibfield  {title} {\bibinfo {title} {Flat optics with designer
  metasurfaces},\ }\href {https://doi.org/10.1038/nmat3839} {\bibfield
  {journal} {\bibinfo  {journal} {Nature materials}\ }\textbf {\bibinfo
  {volume} {13}},\ \bibinfo {pages} {139} (\bibinfo {year} {2014})}\BibitemShut
  {NoStop}%
\bibitem [{\citenamefont {Aieta}\ \emph {et~al.}(2015)\citenamefont {Aieta},
  \citenamefont {Kats}, \citenamefont {Genevet},\ and\ \citenamefont
  {Capasso}}]{aieta2015multiwavelength}%
  \BibitemOpen
  \bibfield  {author} {\bibinfo {author} {\bibfnamefont {F.}~\bibnamefont
  {Aieta}}, \bibinfo {author} {\bibfnamefont {M.~A.}\ \bibnamefont {Kats}},
  \bibinfo {author} {\bibfnamefont {P.}~\bibnamefont {Genevet}},\ and\ \bibinfo
  {author} {\bibfnamefont {F.}~\bibnamefont {Capasso}},\ }\bibfield  {title}
  {\bibinfo {title} {Multiwavelength achromatic metasurfaces by dispersive
  phase compensation},\ }\href {https://doi.org/10.1126/science.aaa2494}
  {\bibfield  {journal} {\bibinfo  {journal} {Science}\ }\textbf {\bibinfo
  {volume} {347}},\ \bibinfo {pages} {1342} (\bibinfo {year}
  {2015})}\BibitemShut {NoStop}%
\bibitem [{\citenamefont {Khorasaninejad}\ \emph {et~al.}(2015)\citenamefont
  {Khorasaninejad}, \citenamefont {Aieta}, \citenamefont {Kanhaiya},
  \citenamefont {Kats}, \citenamefont {Genevet}, \citenamefont {Rousso},\ and\
  \citenamefont {Capasso}}]{khorasaninejad2015achromatic}%
  \BibitemOpen
  \bibfield  {author} {\bibinfo {author} {\bibfnamefont {M.}~\bibnamefont
  {Khorasaninejad}}, \bibinfo {author} {\bibfnamefont {F.}~\bibnamefont
  {Aieta}}, \bibinfo {author} {\bibfnamefont {P.}~\bibnamefont {Kanhaiya}},
  \bibinfo {author} {\bibfnamefont {M.~A.}\ \bibnamefont {Kats}}, \bibinfo
  {author} {\bibfnamefont {P.}~\bibnamefont {Genevet}}, \bibinfo {author}
  {\bibfnamefont {D.}~\bibnamefont {Rousso}},\ and\ \bibinfo {author}
  {\bibfnamefont {F.}~\bibnamefont {Capasso}},\ }\bibfield  {title} {\bibinfo
  {title} {Achromatic metasurface lens at telecommunication wavelengths},\
  }\href {https://doi.org/10.1021/acs.nanolett.5b01727} {\bibfield  {journal}
  {\bibinfo  {journal} {Nano letters}\ }\textbf {\bibinfo {volume} {15}},\
  \bibinfo {pages} {5358} (\bibinfo {year} {2015})}\BibitemShut {NoStop}%
\bibitem [{\citenamefont {Avayu}\ \emph {et~al.}(2017)\citenamefont {Avayu},
  \citenamefont {Almeida}, \citenamefont {Prior},\ and\ \citenamefont
  {Ellenbogen}}]{avayu2017composite}%
  \BibitemOpen
  \bibfield  {author} {\bibinfo {author} {\bibfnamefont {O.}~\bibnamefont
  {Avayu}}, \bibinfo {author} {\bibfnamefont {E.}~\bibnamefont {Almeida}},
  \bibinfo {author} {\bibfnamefont {Y.}~\bibnamefont {Prior}},\ and\ \bibinfo
  {author} {\bibfnamefont {T.}~\bibnamefont {Ellenbogen}},\ }\bibfield  {title}
  {\bibinfo {title} {Composite functional metasurfaces for multispectral
  achromatic optics},\ }\href {https://doi.org/10.1038/ncomms14992} {\bibfield
  {journal} {\bibinfo  {journal} {Nature communications}\ }\textbf {\bibinfo
  {volume} {8}},\ \bibinfo {pages} {14992} (\bibinfo {year}
  {2017})}\BibitemShut {NoStop}%
\bibitem [{\citenamefont {Shrestha}\ \emph {et~al.}(2018)\citenamefont
  {Shrestha}, \citenamefont {Overvig}, \citenamefont {Lu}, \citenamefont
  {Stein},\ and\ \citenamefont {Yu}}]{shrestha2018broadband}%
  \BibitemOpen
  \bibfield  {author} {\bibinfo {author} {\bibfnamefont {S.}~\bibnamefont
  {Shrestha}}, \bibinfo {author} {\bibfnamefont {A.~C.}\ \bibnamefont
  {Overvig}}, \bibinfo {author} {\bibfnamefont {M.}~\bibnamefont {Lu}},
  \bibinfo {author} {\bibfnamefont {A.}~\bibnamefont {Stein}},\ and\ \bibinfo
  {author} {\bibfnamefont {N.}~\bibnamefont {Yu}},\ }\bibfield  {title}
  {\bibinfo {title} {Broadband achromatic dielectric metalenses},\ }\href
  {https://doi.org/10.1038/s41377-018-0078-x} {\bibfield  {journal} {\bibinfo
  {journal} {Light: Science \& Applications}\ }\textbf {\bibinfo {volume}
  {7}},\ \bibinfo {pages} {85} (\bibinfo {year} {2018})}\BibitemShut {NoStop}%
\bibitem [{\citenamefont {Chen}\ \emph {et~al.}(2017)\citenamefont {Chen},
  \citenamefont {Wu}, \citenamefont {Su}, \citenamefont {Lai}, \citenamefont
  {Chu}, \citenamefont {Lee}, \citenamefont {Chen}, \citenamefont {Chen},
  \citenamefont {Lan}, \citenamefont {Kuan},\ and\ \citenamefont
  {Tsai}}]{chen2017gan}%
  \BibitemOpen
  \bibfield  {author} {\bibinfo {author} {\bibfnamefont {B.~H.}\ \bibnamefont
  {Chen}}, \bibinfo {author} {\bibfnamefont {P.~C.}\ \bibnamefont {Wu}},
  \bibinfo {author} {\bibfnamefont {V.-C.}\ \bibnamefont {Su}}, \bibinfo
  {author} {\bibfnamefont {Y.-C.}\ \bibnamefont {Lai}}, \bibinfo {author}
  {\bibfnamefont {C.~H.}\ \bibnamefont {Chu}}, \bibinfo {author} {\bibfnamefont
  {I.~C.}\ \bibnamefont {Lee}}, \bibinfo {author} {\bibfnamefont {J.-W.}\
  \bibnamefont {Chen}}, \bibinfo {author} {\bibfnamefont {Y.~H.}\ \bibnamefont
  {Chen}}, \bibinfo {author} {\bibfnamefont {Y.-C.}\ \bibnamefont {Lan}},
  \bibinfo {author} {\bibfnamefont {C.-H.}\ \bibnamefont {Kuan}},\ and\
  \bibinfo {author} {\bibfnamefont {D.~P.}\ \bibnamefont {Tsai}},\ }\bibfield
  {title} {\bibinfo {title} {Ga{N} metalens for pixel-level full-color routing
  at visible light},\ }\href {https://doi.org/10.1021/acs.nanolett.7b03135}
  {\bibfield  {journal} {\bibinfo  {journal} {Nano letters}\ }\textbf {\bibinfo
  {volume} {17}},\ \bibinfo {pages} {6345} (\bibinfo {year}
  {2017})}\BibitemShut {NoStop}%
\bibitem [{\citenamefont {Paniagua-Dominguez}\ \emph
  {et~al.}(2018)\citenamefont {Paniagua-Dominguez}, \citenamefont {Yu},
  \citenamefont {Khaidarov}, \citenamefont {Choi}, \citenamefont {Leong},
  \citenamefont {Bakker}, \citenamefont {Liang}, \citenamefont {Fu},
  \citenamefont {Valuckas}, \citenamefont {Krivitsky},\ and\ \citenamefont
  {Kuznetsov}}]{paniagua2018metalens}%
  \BibitemOpen
  \bibfield  {author} {\bibinfo {author} {\bibfnamefont {R.}~\bibnamefont
  {Paniagua-Dominguez}}, \bibinfo {author} {\bibfnamefont {Y.~F.}\ \bibnamefont
  {Yu}}, \bibinfo {author} {\bibfnamefont {E.}~\bibnamefont {Khaidarov}},
  \bibinfo {author} {\bibfnamefont {S.}~\bibnamefont {Choi}}, \bibinfo {author}
  {\bibfnamefont {V.}~\bibnamefont {Leong}}, \bibinfo {author} {\bibfnamefont
  {R.~M.}\ \bibnamefont {Bakker}}, \bibinfo {author} {\bibfnamefont
  {X.}~\bibnamefont {Liang}}, \bibinfo {author} {\bibfnamefont {Y.~H.}\
  \bibnamefont {Fu}}, \bibinfo {author} {\bibfnamefont {V.}~\bibnamefont
  {Valuckas}}, \bibinfo {author} {\bibfnamefont {L.~A.}\ \bibnamefont
  {Krivitsky}},\ and\ \bibinfo {author} {\bibfnamefont {A.~I.}\ \bibnamefont
  {Kuznetsov}},\ }\bibfield  {title} {\bibinfo {title} {A metalens with a
  near-unity numerical aperture},\ }\href
  {https://doi.org/10.1021/acs.nanolett.8b00368} {\bibfield  {journal}
  {\bibinfo  {journal} {Nano letters}\ }\textbf {\bibinfo {volume} {18}},\
  \bibinfo {pages} {2124} (\bibinfo {year} {2018})}\BibitemShut {NoStop}%
\bibitem [{\citenamefont {Chen}\ \emph {et~al.}(2018)\citenamefont {Chen},
  \citenamefont {Zhu}, \citenamefont {Sanjeev}, \citenamefont {Khorasaninejad},
  \citenamefont {Shi}, \citenamefont {Lee},\ and\ \citenamefont
  {Capasso}}]{chen2018broadband}%
  \BibitemOpen
  \bibfield  {author} {\bibinfo {author} {\bibfnamefont {W.~T.}\ \bibnamefont
  {Chen}}, \bibinfo {author} {\bibfnamefont {A.~Y.}\ \bibnamefont {Zhu}},
  \bibinfo {author} {\bibfnamefont {V.}~\bibnamefont {Sanjeev}}, \bibinfo
  {author} {\bibfnamefont {M.}~\bibnamefont {Khorasaninejad}}, \bibinfo
  {author} {\bibfnamefont {Z.}~\bibnamefont {Shi}}, \bibinfo {author}
  {\bibfnamefont {E.}~\bibnamefont {Lee}},\ and\ \bibinfo {author}
  {\bibfnamefont {F.}~\bibnamefont {Capasso}},\ }\bibfield  {title} {\bibinfo
  {title} {A broadband achromatic metalens for focusing and imaging in the
  visible},\ }\href {https://doi.org/10.1038/s41565-017-0034-6} {\bibfield
  {journal} {\bibinfo  {journal} {Nature nanotechnology}\ }\textbf {\bibinfo
  {volume} {13}},\ \bibinfo {pages} {220} (\bibinfo {year} {2018})}\BibitemShut
  {NoStop}%
\bibitem [{\citenamefont {Wang}\ \emph {et~al.}(2018)\citenamefont {Wang},
  \citenamefont {Wu}, \citenamefont {Su}, \citenamefont {Lai}, \citenamefont
  {Chen}, \citenamefont {Kuo}, \citenamefont {Chen}, \citenamefont {Chen},
  \citenamefont {Huang}, \citenamefont {Wang}, \citenamefont {Lin},
  \citenamefont {Kuan}, \citenamefont {Li}, \citenamefont {Wang}, \citenamefont
  {Zhu},\ and\ \citenamefont {Tsai}}]{wang2018broadband}%
  \BibitemOpen
  \bibfield  {author} {\bibinfo {author} {\bibfnamefont {S.}~\bibnamefont
  {Wang}}, \bibinfo {author} {\bibfnamefont {P.~C.}\ \bibnamefont {Wu}},
  \bibinfo {author} {\bibfnamefont {V.-C.}\ \bibnamefont {Su}}, \bibinfo
  {author} {\bibfnamefont {Y.-C.}\ \bibnamefont {Lai}}, \bibinfo {author}
  {\bibfnamefont {M.-K.}\ \bibnamefont {Chen}}, \bibinfo {author}
  {\bibfnamefont {H.~Y.}\ \bibnamefont {Kuo}}, \bibinfo {author} {\bibfnamefont
  {B.~H.}\ \bibnamefont {Chen}}, \bibinfo {author} {\bibfnamefont {Y.~H.}\
  \bibnamefont {Chen}}, \bibinfo {author} {\bibfnamefont {T.-T.}\ \bibnamefont
  {Huang}}, \bibinfo {author} {\bibfnamefont {J.-H.}\ \bibnamefont {Wang}},
  \bibinfo {author} {\bibfnamefont {R.-M.}\ \bibnamefont {Lin}}, \bibinfo
  {author} {\bibfnamefont {C.-H.}\ \bibnamefont {Kuan}}, \bibinfo {author}
  {\bibfnamefont {T.}~\bibnamefont {Li}}, \bibinfo {author} {\bibfnamefont
  {Z.}~\bibnamefont {Wang}}, \bibinfo {author} {\bibfnamefont {S.}~\bibnamefont
  {Zhu}},\ and\ \bibinfo {author} {\bibfnamefont {D.~P.}\ \bibnamefont
  {Tsai}},\ }\bibfield  {title} {\bibinfo {title} {A broadband achromatic
  metalens in the visible},\ }\href {https://doi.org/10.1038/s41565-017-0052-4}
  {\bibfield  {journal} {\bibinfo  {journal} {Nature nanotechnology}\ }\textbf
  {\bibinfo {volume} {13}},\ \bibinfo {pages} {227} (\bibinfo {year}
  {2018})}\BibitemShut {NoStop}%
\bibitem [{\citenamefont {Bends{\o}e}\ and\ \citenamefont
  {Sigmund}(1999)}]{bendsoe1999material}%
  \BibitemOpen
  \bibfield  {author} {\bibinfo {author} {\bibfnamefont {M.~P.}\ \bibnamefont
  {Bends{\o}e}}\ and\ \bibinfo {author} {\bibfnamefont {O.}~\bibnamefont
  {Sigmund}},\ }\bibfield  {title} {\bibinfo {title} {Material interpolation
  schemes in topology optimization},\ }\href
  {https://doi.org/10.1007/s004190050248} {\bibfield  {journal} {\bibinfo
  {journal} {Archive of applied mechanics}\ }\textbf {\bibinfo {volume} {69}},\
  \bibinfo {pages} {635} (\bibinfo {year} {1999})}\BibitemShut {NoStop}%
\bibitem [{\citenamefont {Bends{\o}e}(2001)}]{bendsoe2001topology}%
  \BibitemOpen
  \bibfield  {author} {\bibinfo {author} {\bibfnamefont {M.~P.}\ \bibnamefont
  {Bends{\o}e}},\ }\bibfield  {title} {\bibinfo {title} {Topology
  optimization},\ }in\ \href@noop {} {\emph {\bibinfo {booktitle} {Encyclopedia
  of Optimization}}}\ (\bibinfo  {publisher} {Springer},\ \bibinfo {year}
  {2001})\ pp.\ \bibinfo {pages} {2636--2638}\BibitemShut {NoStop}%
\bibitem [{\citenamefont {Neves}\ \emph {et~al.}(2002)\citenamefont {Neves},
  \citenamefont {Sigmund},\ and\ \citenamefont
  {Bends{\o}e}}]{neves2002topology}%
  \BibitemOpen
  \bibfield  {author} {\bibinfo {author} {\bibfnamefont {M.~M.}\ \bibnamefont
  {Neves}}, \bibinfo {author} {\bibfnamefont {O.}~\bibnamefont {Sigmund}},\
  and\ \bibinfo {author} {\bibfnamefont {M.~P.}\ \bibnamefont {Bends{\o}e}},\
  }\bibfield  {title} {\bibinfo {title} {Topology optimization of periodic
  microstructures with a penalization of highly localized buckling modes},\
  }\href {https://doi.org/10.1002/nme.449} {\bibfield  {journal} {\bibinfo
  {journal} {International Journal for Numerical Methods in Engineering}\
  }\textbf {\bibinfo {volume} {54}},\ \bibinfo {pages} {809} (\bibinfo {year}
  {2002})}\BibitemShut {NoStop}%
\bibitem [{\citenamefont {Molesky}\ \emph {et~al.}(2018)\citenamefont
  {Molesky}, \citenamefont {Lin}, \citenamefont {Piggott}, \citenamefont {Jin},
  \citenamefont {Vuckovi{\'c}},\ and\ \citenamefont
  {Rodriguez}}]{molesky2018inverse}%
  \BibitemOpen
  \bibfield  {author} {\bibinfo {author} {\bibfnamefont {S.}~\bibnamefont
  {Molesky}}, \bibinfo {author} {\bibfnamefont {Z.}~\bibnamefont {Lin}},
  \bibinfo {author} {\bibfnamefont {A.~Y.}\ \bibnamefont {Piggott}}, \bibinfo
  {author} {\bibfnamefont {W.}~\bibnamefont {Jin}}, \bibinfo {author}
  {\bibfnamefont {J.}~\bibnamefont {Vuckovi{\'c}}},\ and\ \bibinfo {author}
  {\bibfnamefont {A.~W.}\ \bibnamefont {Rodriguez}},\ }\bibfield  {title}
  {\bibinfo {title} {Inverse design in nanophotonics},\ }\href
  {https://doi.org/10.1038/s41566-018-0246-9} {\bibfield  {journal} {\bibinfo
  {journal} {Nature Photonics}\ }\textbf {\bibinfo {volume} {12}},\ \bibinfo
  {pages} {659} (\bibinfo {year} {2018})}\BibitemShut {NoStop}%
\bibitem [{\citenamefont {Miller}(2012)}]{miller2012photonic}%
  \BibitemOpen
  \bibfield  {author} {\bibinfo {author} {\bibfnamefont {O.~D.}\ \bibnamefont
  {Miller}},\ }\emph {\bibinfo {title} {Photonic Design: From Fundamental Solar
  Cell Physics to Computational Inverse Design}},\ \href@noop {} {Ph.D.
  thesis},\ \bibinfo  {school} {University of California, Berkeley} (\bibinfo
  {year} {2012})\BibitemShut {NoStop}%
\bibitem [{\citenamefont {Pestourie}\ \emph {et~al.}(2018)\citenamefont
  {Pestourie}, \citenamefont {P{\'e}rez-Arancibia}, \citenamefont {Lin},
  \citenamefont {Shin}, \citenamefont {Capasso},\ and\ \citenamefont
  {Johnson}}]{pestourie2018inverse}%
  \BibitemOpen
  \bibfield  {author} {\bibinfo {author} {\bibfnamefont {R.}~\bibnamefont
  {Pestourie}}, \bibinfo {author} {\bibfnamefont {C.}~\bibnamefont
  {P{\'e}rez-Arancibia}}, \bibinfo {author} {\bibfnamefont {Z.}~\bibnamefont
  {Lin}}, \bibinfo {author} {\bibfnamefont {W.}~\bibnamefont {Shin}}, \bibinfo
  {author} {\bibfnamefont {F.}~\bibnamefont {Capasso}},\ and\ \bibinfo {author}
  {\bibfnamefont {S.~G.}\ \bibnamefont {Johnson}},\ }\bibfield  {title}
  {\bibinfo {title} {Inverse design of large-area metasurfaces},\ }\href
  {https://doi.org/10.1364/OE.26.033732} {\bibfield  {journal} {\bibinfo
  {journal} {Optics express}\ }\textbf {\bibinfo {volume} {26}},\ \bibinfo
  {pages} {33732} (\bibinfo {year} {2018})}\BibitemShut {NoStop}%
\bibitem [{\citenamefont {Camayd-Mu{\~n}oz}\ and\ \citenamefont
  {Faraon}(2018)}]{camayd2018scaling}%
  \BibitemOpen
  \bibfield  {author} {\bibinfo {author} {\bibfnamefont {P.}~\bibnamefont
  {Camayd-Mu{\~n}oz}}\ and\ \bibinfo {author} {\bibfnamefont {A.}~\bibnamefont
  {Faraon}},\ }\bibfield  {title} {\bibinfo {title} {Scaling laws for
  inverse-designed metadevices},\ }in\ \href
  {https://doi.org/10.1364/CLEO_QELS.2018.FF3C.7} {\emph {\bibinfo {booktitle}
  {Conference on Lasers and Electro-Optics}}}\ (\bibinfo {organization}
  {Optical Society of America},\ \bibinfo {year} {2018})\ pp.\ \bibinfo {pages}
  {FF3C--7}\BibitemShut {NoStop}%
\bibitem [{\citenamefont {Piggott}\ \emph {et~al.}(2015)\citenamefont
  {Piggott}, \citenamefont {Lu}, \citenamefont {Lagoudakis}, \citenamefont
  {Petykiewicz}, \citenamefont {Babinec},\ and\ \citenamefont
  {Vu{\v{c}}kovi{\'c}}}]{piggott2015inverse}%
  \BibitemOpen
  \bibfield  {author} {\bibinfo {author} {\bibfnamefont {A.~Y.}\ \bibnamefont
  {Piggott}}, \bibinfo {author} {\bibfnamefont {J.}~\bibnamefont {Lu}},
  \bibinfo {author} {\bibfnamefont {K.~G.}\ \bibnamefont {Lagoudakis}},
  \bibinfo {author} {\bibfnamefont {J.}~\bibnamefont {Petykiewicz}}, \bibinfo
  {author} {\bibfnamefont {T.~M.}\ \bibnamefont {Babinec}},\ and\ \bibinfo
  {author} {\bibfnamefont {J.}~\bibnamefont {Vu{\v{c}}kovi{\'c}}},\ }\bibfield
  {title} {\bibinfo {title} {Inverse design and demonstration of a compact and
  broadband on-chip wavelength demultiplexer},\ }\href
  {https://doi.org/10.1038/nphoton.2015.69} {\bibfield  {journal} {\bibinfo
  {journal} {Nature Photonics}\ }\textbf {\bibinfo {volume} {9}},\ \bibinfo
  {pages} {374} (\bibinfo {year} {2015})}\BibitemShut {NoStop}%
\bibitem [{\citenamefont {Jensen}\ and\ \citenamefont
  {Sigmund}(2011)}]{jensen2011topology}%
  \BibitemOpen
  \bibfield  {author} {\bibinfo {author} {\bibfnamefont {J.~S.}\ \bibnamefont
  {Jensen}}\ and\ \bibinfo {author} {\bibfnamefont {O.}~\bibnamefont
  {Sigmund}},\ }\bibfield  {title} {\bibinfo {title} {Topology optimization for
  nano-photonics},\ }\href {https://doi.org/10.1002/lpor.201000014} {\bibfield
  {journal} {\bibinfo  {journal} {Laser \& Photonics Reviews}\ }\textbf
  {\bibinfo {volume} {5}},\ \bibinfo {pages} {308} (\bibinfo {year}
  {2011})}\BibitemShut {NoStop}%
\bibitem [{\citenamefont {Yang}\ \emph {et~al.}(2009)\citenamefont {Yang},
  \citenamefont {Lavrinenko}, \citenamefont {Hvam},\ and\ \citenamefont
  {Sigmund}}]{yang2009design}%
  \BibitemOpen
  \bibfield  {author} {\bibinfo {author} {\bibfnamefont {L.}~\bibnamefont
  {Yang}}, \bibinfo {author} {\bibfnamefont {A.~V.}\ \bibnamefont
  {Lavrinenko}}, \bibinfo {author} {\bibfnamefont {J.~M.}\ \bibnamefont
  {Hvam}},\ and\ \bibinfo {author} {\bibfnamefont {O.}~\bibnamefont
  {Sigmund}},\ }\bibfield  {title} {\bibinfo {title} {Design of one-dimensional
  optical pulse-shaping filters by time-domain topology optimization},\ }\href
  {https://doi.org/10.1063/1.3278595} {\bibfield  {journal} {\bibinfo
  {journal} {Applied Physics Letters}\ }\textbf {\bibinfo {volume} {95}},\
  \bibinfo {pages} {261101} (\bibinfo {year} {2009})}\BibitemShut {NoStop}%
\bibitem [{\citenamefont {Lin}\ \emph {et~al.}(2018)\citenamefont {Lin},
  \citenamefont {Groever}, \citenamefont {Capasso}, \citenamefont {Rodriguez},\
  and\ \citenamefont {Lon{\v{c}}ar}}]{lin2018topology}%
  \BibitemOpen
  \bibfield  {author} {\bibinfo {author} {\bibfnamefont {Z.}~\bibnamefont
  {Lin}}, \bibinfo {author} {\bibfnamefont {B.}~\bibnamefont {Groever}},
  \bibinfo {author} {\bibfnamefont {F.}~\bibnamefont {Capasso}}, \bibinfo
  {author} {\bibfnamefont {A.~W.}\ \bibnamefont {Rodriguez}},\ and\ \bibinfo
  {author} {\bibfnamefont {M.}~\bibnamefont {Lon{\v{c}}ar}},\ }\bibfield
  {title} {\bibinfo {title} {Topology-optimized multilayered metaoptics},\
  }\href {https://doi.org/10.1103/PhysRevApplied.9.044030} {\bibfield
  {journal} {\bibinfo  {journal} {Physical Review Applied}\ }\textbf {\bibinfo
  {volume} {9}},\ \bibinfo {pages} {044030} (\bibinfo {year}
  {2018})}\BibitemShut {NoStop}%
\bibitem [{\citenamefont {Mohammad}\ \emph {et~al.}(2018)\citenamefont
  {Mohammad}, \citenamefont {Meem}, \citenamefont {Shen}, \citenamefont
  {Wang},\ and\ \citenamefont {Menon}}]{mohammad2018broadband}%
  \BibitemOpen
  \bibfield  {author} {\bibinfo {author} {\bibfnamefont {N.}~\bibnamefont
  {Mohammad}}, \bibinfo {author} {\bibfnamefont {M.}~\bibnamefont {Meem}},
  \bibinfo {author} {\bibfnamefont {B.}~\bibnamefont {Shen}}, \bibinfo {author}
  {\bibfnamefont {P.}~\bibnamefont {Wang}},\ and\ \bibinfo {author}
  {\bibfnamefont {R.}~\bibnamefont {Menon}},\ }\bibfield  {title} {\bibinfo
  {title} {Broadband imaging with one planar diffractive lens},\ }\href
  {https://doi.org/10.1038/s41598-018-21169-4} {\bibfield  {journal} {\bibinfo
  {journal} {Scientific reports}\ }\textbf {\bibinfo {volume} {8}},\ \bibinfo
  {pages} {2799} (\bibinfo {year} {2018})}\BibitemShut {NoStop}%
\bibitem [{\citenamefont {Zheng}\ \emph {et~al.}(2015)\citenamefont {Zheng},
  \citenamefont {M{\"u}hlenbernd}, \citenamefont {Kenney}, \citenamefont {Li},
  \citenamefont {Zentgraf},\ and\ \citenamefont
  {Zhang}}]{zheng2015metasurface}%
  \BibitemOpen
  \bibfield  {author} {\bibinfo {author} {\bibfnamefont {G.}~\bibnamefont
  {Zheng}}, \bibinfo {author} {\bibfnamefont {H.}~\bibnamefont
  {M{\"u}hlenbernd}}, \bibinfo {author} {\bibfnamefont {M.}~\bibnamefont
  {Kenney}}, \bibinfo {author} {\bibfnamefont {G.}~\bibnamefont {Li}}, \bibinfo
  {author} {\bibfnamefont {T.}~\bibnamefont {Zentgraf}},\ and\ \bibinfo
  {author} {\bibfnamefont {S.}~\bibnamefont {Zhang}},\ }\bibfield  {title}
  {\bibinfo {title} {Metasurface holograms reaching 80\% efficiency},\ }\href
  {https://doi.org/10.1038/nnano.2015.2} {\bibfield  {journal} {\bibinfo
  {journal} {Nature nanotechnology}\ }\textbf {\bibinfo {volume} {10}},\
  \bibinfo {pages} {308} (\bibinfo {year} {2015})}\BibitemShut {NoStop}%
\bibitem [{\citenamefont {Ni}\ \emph {et~al.}(2013)\citenamefont {Ni},
  \citenamefont {Kildishev},\ and\ \citenamefont
  {Shalaev}}]{ni2013metasurface}%
  \BibitemOpen
  \bibfield  {author} {\bibinfo {author} {\bibfnamefont {X.}~\bibnamefont
  {Ni}}, \bibinfo {author} {\bibfnamefont {A.~V.}\ \bibnamefont {Kildishev}},\
  and\ \bibinfo {author} {\bibfnamefont {V.~M.}\ \bibnamefont {Shalaev}},\
  }\bibfield  {title} {\bibinfo {title} {Metasurface holograms for visible
  light},\ }\href {https://doi.org/10.1038/ncomms3807} {\bibfield  {journal}
  {\bibinfo  {journal} {Nature communications}\ }\textbf {\bibinfo {volume}
  {4}},\ \bibinfo {pages} {2807} (\bibinfo {year} {2013})}\BibitemShut
  {NoStop}%
\bibitem [{\citenamefont {Zhao}\ \emph {et~al.}(2016)\citenamefont {Zhao},
  \citenamefont {Liu}, \citenamefont {Jiang}, \citenamefont {Song},
  \citenamefont {Pei},\ and\ \citenamefont {Jiang}}]{zhao2016full}%
  \BibitemOpen
  \bibfield  {author} {\bibinfo {author} {\bibfnamefont {W.}~\bibnamefont
  {Zhao}}, \bibinfo {author} {\bibfnamefont {B.}~\bibnamefont {Liu}}, \bibinfo
  {author} {\bibfnamefont {H.}~\bibnamefont {Jiang}}, \bibinfo {author}
  {\bibfnamefont {J.}~\bibnamefont {Song}}, \bibinfo {author} {\bibfnamefont
  {Y.}~\bibnamefont {Pei}},\ and\ \bibinfo {author} {\bibfnamefont
  {Y.}~\bibnamefont {Jiang}},\ }\bibfield  {title} {\bibinfo {title}
  {Full-color hologram using spatial multiplexing of dielectric metasurface},\
  }\href {https://doi.org/10.1364/OL.41.000147} {\bibfield  {journal} {\bibinfo
   {journal} {Optics letters}\ }\textbf {\bibinfo {volume} {41}},\ \bibinfo
  {pages} {147} (\bibinfo {year} {2016})}\BibitemShut {NoStop}%
\bibitem [{\citenamefont {Arbabi}\ \emph {et~al.}(2017)\citenamefont {Arbabi},
  \citenamefont {Arbabi}, \citenamefont {Horie}, \citenamefont {Kamali},\ and\
  \citenamefont {Faraon}}]{arbabi2017planar}%
  \BibitemOpen
  \bibfield  {author} {\bibinfo {author} {\bibfnamefont {A.}~\bibnamefont
  {Arbabi}}, \bibinfo {author} {\bibfnamefont {E.}~\bibnamefont {Arbabi}},
  \bibinfo {author} {\bibfnamefont {Y.}~\bibnamefont {Horie}}, \bibinfo
  {author} {\bibfnamefont {S.~M.}\ \bibnamefont {Kamali}},\ and\ \bibinfo
  {author} {\bibfnamefont {A.}~\bibnamefont {Faraon}},\ }\bibfield  {title}
  {\bibinfo {title} {Planar metasurface retroreflector},\ }\href
  {https://doi.org/10.1038/nphoton.2017.96} {\bibfield  {journal} {\bibinfo
  {journal} {Nature Photonics}\ }\textbf {\bibinfo {volume} {11}},\ \bibinfo
  {pages} {415} (\bibinfo {year} {2017})}\BibitemShut {NoStop}%
\bibitem [{\citenamefont {Khorasaninejad}\ \emph
  {et~al.}(2016{\natexlab{a}})\citenamefont {Khorasaninejad}, \citenamefont
  {Chen}, \citenamefont {Devlin}, \citenamefont {Oh}, \citenamefont {Zhu},\
  and\ \citenamefont {Capasso}}]{khorasaninejad2016metalenses}%
  \BibitemOpen
  \bibfield  {author} {\bibinfo {author} {\bibfnamefont {M.}~\bibnamefont
  {Khorasaninejad}}, \bibinfo {author} {\bibfnamefont {W.~T.}\ \bibnamefont
  {Chen}}, \bibinfo {author} {\bibfnamefont {R.~C.}\ \bibnamefont {Devlin}},
  \bibinfo {author} {\bibfnamefont {J.}~\bibnamefont {Oh}}, \bibinfo {author}
  {\bibfnamefont {A.~Y.}\ \bibnamefont {Zhu}},\ and\ \bibinfo {author}
  {\bibfnamefont {F.}~\bibnamefont {Capasso}},\ }\bibfield  {title} {\bibinfo
  {title} {Metalenses at visible wavelengths: Diffraction-limited focusing and
  subwavelength resolution imaging},\ }\href
  {https://doi.org/10.1126/science.aaf6644} {\bibfield  {journal} {\bibinfo
  {journal} {Science}\ }\textbf {\bibinfo {volume} {352}},\ \bibinfo {pages}
  {1190} (\bibinfo {year} {2016}{\natexlab{a}})}\BibitemShut {NoStop}%
\bibitem [{\citenamefont {Kamali}\ \emph {et~al.}(2016)\citenamefont {Kamali},
  \citenamefont {Arbabi}, \citenamefont {Arbabi}, \citenamefont {Horie},\ and\
  \citenamefont {Faraon}}]{kamali2016highly}%
  \BibitemOpen
  \bibfield  {author} {\bibinfo {author} {\bibfnamefont {S.~M.}\ \bibnamefont
  {Kamali}}, \bibinfo {author} {\bibfnamefont {E.}~\bibnamefont {Arbabi}},
  \bibinfo {author} {\bibfnamefont {A.}~\bibnamefont {Arbabi}}, \bibinfo
  {author} {\bibfnamefont {Y.}~\bibnamefont {Horie}},\ and\ \bibinfo {author}
  {\bibfnamefont {A.}~\bibnamefont {Faraon}},\ }\bibfield  {title} {\bibinfo
  {title} {Highly tunable elastic dielectric metasurface lenses},\ }\href
  {https://doi.org/10.1002/lpor.201600144} {\bibfield  {journal} {\bibinfo
  {journal} {Laser \& Photonics Reviews}\ }\textbf {\bibinfo {volume} {10}},\
  \bibinfo {pages} {1002} (\bibinfo {year} {2016})}\BibitemShut {NoStop}%
\bibitem [{\citenamefont {Khorasaninejad}\ \emph
  {et~al.}(2016{\natexlab{b}})\citenamefont {Khorasaninejad}, \citenamefont
  {Zhu}, \citenamefont {Roques-Carmes}, \citenamefont {Chen}, \citenamefont
  {Oh}, \citenamefont {Mishra}, \citenamefont {Devlin},\ and\ \citenamefont
  {Capasso}}]{khorasaninejad2016polarization}%
  \BibitemOpen
  \bibfield  {author} {\bibinfo {author} {\bibfnamefont {M.}~\bibnamefont
  {Khorasaninejad}}, \bibinfo {author} {\bibfnamefont {A.~Y.}\ \bibnamefont
  {Zhu}}, \bibinfo {author} {\bibfnamefont {C.}~\bibnamefont {Roques-Carmes}},
  \bibinfo {author} {\bibfnamefont {W.~T.}\ \bibnamefont {Chen}}, \bibinfo
  {author} {\bibfnamefont {J.}~\bibnamefont {Oh}}, \bibinfo {author}
  {\bibfnamefont {I.}~\bibnamefont {Mishra}}, \bibinfo {author} {\bibfnamefont
  {R.~C.}\ \bibnamefont {Devlin}},\ and\ \bibinfo {author} {\bibfnamefont
  {F.}~\bibnamefont {Capasso}},\ }\bibfield  {title} {\bibinfo {title}
  {Polarization-insensitive metalenses at visible wavelengths},\ }\href
  {https://doi.org/10.1021/acs.nanolett.6b03626} {\bibfield  {journal}
  {\bibinfo  {journal} {Nano letters}\ }\textbf {\bibinfo {volume} {16}},\
  \bibinfo {pages} {7229} (\bibinfo {year} {2016}{\natexlab{b}})}\BibitemShut
  {NoStop}%
\bibitem [{\citenamefont {Pontryagin}\ \emph {et~al.}(1962)\citenamefont
  {Pontryagin}, \citenamefont {Boltyanskii}, \citenamefont {Gamkrelidze},\ and\
  \citenamefont {Mishechenko}}]{Pontryagin1962}%
  \BibitemOpen
  \bibfield  {author} {\bibinfo {author} {\bibfnamefont {L.~S.}\ \bibnamefont
  {Pontryagin}}, \bibinfo {author} {\bibfnamefont {V.~G.}\ \bibnamefont
  {Boltyanskii}}, \bibinfo {author} {\bibfnamefont {R.~V.}\ \bibnamefont
  {Gamkrelidze}},\ and\ \bibinfo {author} {\bibfnamefont {E.~F.}\ \bibnamefont
  {Mishechenko}},\ }\href@noop {} {\emph {\bibinfo {title} {{The Mathematical
  Theory of Optimal Processes}}}}\ (\bibinfo  {publisher} {John Wiley {\&}
  Sons},\ \bibinfo {year} {1962})\BibitemShut {NoStop}%
\bibitem [{\citenamefont {C{\'e}a}\ \emph {et~al.}(1973)\citenamefont
  {C{\'e}a}, \citenamefont {Gioan},\ and\ \citenamefont
  {Michel}}]{cea1973quelques}%
  \BibitemOpen
  \bibfield  {author} {\bibinfo {author} {\bibfnamefont {J.}~\bibnamefont
  {C{\'e}a}}, \bibinfo {author} {\bibfnamefont {A.}~\bibnamefont {Gioan}},\
  and\ \bibinfo {author} {\bibfnamefont {J.}~\bibnamefont {Michel}},\
  }\bibfield  {title} {\bibinfo {title} {Quelques r{\'e}sultats sur
  l'identification de domaines},\ }\href {https://doi.org/10.1007/BF02575843}
  {\bibfield  {journal} {\bibinfo  {journal} {Calcolo}\ }\textbf {\bibinfo
  {volume} {10}},\ \bibinfo {pages} {207} (\bibinfo {year} {1973})}\BibitemShut
  {NoStop}%
\bibitem [{\citenamefont {Pironneau}(2012)}]{Pironneau2012}%
  \BibitemOpen
  \bibfield  {author} {\bibinfo {author} {\bibfnamefont {O.}~\bibnamefont
  {Pironneau}},\ }\href@noop {} {\emph {\bibinfo {title} {{Optimal Shape Design
  for Elliptic Systems}}}}\ (\bibinfo  {publisher} {Springer-Verlag},\ \bibinfo
  {year} {2012})\BibitemShut {NoStop}%
\bibitem [{\citenamefont {Giles}\ and\ \citenamefont
  {Pierce}(2000)}]{Giles2000}%
  \BibitemOpen
  \bibfield  {author} {\bibinfo {author} {\bibfnamefont {M.~B.}\ \bibnamefont
  {Giles}}\ and\ \bibinfo {author} {\bibfnamefont {N.~A.}\ \bibnamefont
  {Pierce}},\ }\bibfield  {title} {\bibinfo {title} {{An Introduction to the
  Adjoint Approach to Design}},\ }\href
  {https://doi.org/10.1023/A:1011430410075} {\bibfield  {journal} {\bibinfo
  {journal} {Flow, Turbul. Combust.}\ }\textbf {\bibinfo {volume} {65}},\
  \bibinfo {pages} {393} (\bibinfo {year} {2000})}\BibitemShut {NoStop}%
\bibitem [{\citenamefont {Director}\ and\ \citenamefont
  {Rohrer}(1969)}]{Director1969}%
  \BibitemOpen
  \bibfield  {author} {\bibinfo {author} {\bibfnamefont {S.~W.}\ \bibnamefont
  {Director}}\ and\ \bibinfo {author} {\bibfnamefont {R.~A.}\ \bibnamefont
  {Rohrer}},\ }\bibfield  {title} {\bibinfo {title} {{The Generalized Adjoint
  Network and Network Sensitivities}},\ }\href
  {https://doi.org/10.1109/TCT.1969.1082965} {\bibfield  {journal} {\bibinfo
  {journal} {IEEE Trans. Circuit Theory}\ }\textbf {\bibinfo {volume} {16}},\
  \bibinfo {pages} {318} (\bibinfo {year} {1969})}\BibitemShut {NoStop}%
\bibitem [{\citenamefont {Jameson}(1988)}]{Jameson1988}%
  \BibitemOpen
  \bibfield  {author} {\bibinfo {author} {\bibfnamefont {A.}~\bibnamefont
  {Jameson}},\ }\bibfield  {title} {\bibinfo {title} {{Aerodynamic design via
  control theory}},\ }\href {https://doi.org/10.1007/BF01061285} {\bibfield
  {journal} {\bibinfo  {journal} {J. Sci. Comput.}\ }\textbf {\bibinfo {volume}
  {3}},\ \bibinfo {pages} {233} (\bibinfo {year} {1988})}\BibitemShut {NoStop}%
\bibitem [{\citenamefont {Bendsoe}\ and\ \citenamefont
  {Sigmund}(2013)}]{Bendsoe2013}%
  \BibitemOpen
  \bibfield  {author} {\bibinfo {author} {\bibfnamefont {M.~P.}\ \bibnamefont
  {Bendsoe}}\ and\ \bibinfo {author} {\bibfnamefont {O.}~\bibnamefont
  {Sigmund}},\ }\href {https://doi.org/10.1007/978-3-662-05086-6} {\emph
  {\bibinfo {title} {{Topology optimization: theory, methods, and
  applications}}}}\ (\bibinfo  {publisher} {Springer Science {\&} Business
  Media},\ \bibinfo {year} {2013})\BibitemShut {NoStop}%
\bibitem [{\citenamefont {Demiralp}\ and\ \citenamefont
  {Rabitz}(1993)}]{Demiralp1993}%
  \BibitemOpen
  \bibfield  {author} {\bibinfo {author} {\bibfnamefont {M.}~\bibnamefont
  {Demiralp}}\ and\ \bibinfo {author} {\bibfnamefont {H.}~\bibnamefont
  {Rabitz}},\ }\bibfield  {title} {\bibinfo {title} {{Optimally controlled
  quantum molecular dynamics: A perturbation formulation and the existence of
  multiple solutions}},\ }\href {https://doi.org/10.1103/PhysRevA.47.809}
  {\bibfield  {journal} {\bibinfo  {journal} {Phys. Rev. A}\ }\textbf {\bibinfo
  {volume} {47}},\ \bibinfo {pages} {809} (\bibinfo {year} {1993})}\BibitemShut
  {NoStop}%
\bibitem [{\citenamefont {Rabitz}\ \emph {et~al.}(2004)\citenamefont {Rabitz},
  \citenamefont {Hsieh},\ and\ \citenamefont {Rosenthal}}]{Rabitz2004}%
  \BibitemOpen
  \bibfield  {author} {\bibinfo {author} {\bibfnamefont {H.~A.}\ \bibnamefont
  {Rabitz}}, \bibinfo {author} {\bibfnamefont {M.~M.}\ \bibnamefont {Hsieh}},\
  and\ \bibinfo {author} {\bibfnamefont {C.~M.}\ \bibnamefont {Rosenthal}},\
  }\bibfield  {title} {\bibinfo {title} {{Quantum Optimally Controlled
  Transition Landscapes}},\ }\href {https://doi.org/10.1126/science.1093649}
  {\bibfield  {journal} {\bibinfo  {journal} {Science}\ }\textbf {\bibinfo
  {volume} {303}},\ \bibinfo {pages} {1998} (\bibinfo {year}
  {2004})}\BibitemShut {NoStop}%
\bibitem [{\citenamefont {Werbos}(1994)}]{Werbos1994}%
  \BibitemOpen
  \bibfield  {author} {\bibinfo {author} {\bibfnamefont {P.~J.}\ \bibnamefont
  {Werbos}},\ }\href@noop {} {\emph {\bibinfo {title} {{The Roots of
  Backpropagation}}}}\ (\bibinfo  {publisher} {John Wiley {\&} Sons, Inc.},\
  \bibinfo {year} {1994})\BibitemShut {NoStop}%
\bibitem [{\citenamefont {Rumelhart}\ \emph {et~al.}(1986)\citenamefont
  {Rumelhart}, \citenamefont {Hinton},\ and\ \citenamefont
  {Williams}}]{Rumelhart1986}%
  \BibitemOpen
  \bibfield  {author} {\bibinfo {author} {\bibfnamefont {D.~E.}\ \bibnamefont
  {Rumelhart}}, \bibinfo {author} {\bibfnamefont {G.~E.}\ \bibnamefont
  {Hinton}},\ and\ \bibinfo {author} {\bibfnamefont {R.~J.}\ \bibnamefont
  {Williams}},\ }\bibfield  {title} {\bibinfo {title} {{Learning
  representations by back-propagating errors}},\ }\href
  {https://doi.org/10.1038/323533a0} {\bibfield  {journal} {\bibinfo  {journal}
  {Nature}\ }\textbf {\bibinfo {volume} {323}},\ \bibinfo {pages} {533}
  (\bibinfo {year} {1986})}\BibitemShut {NoStop}%
\bibitem [{\citenamefont {LeCun}\ \emph {et~al.}(1989)\citenamefont {LeCun},
  \citenamefont {Boser}, \citenamefont {Denker}, \citenamefont {Henderson},
  \citenamefont {Howard}, \citenamefont {Hubbard},\ and\ \citenamefont
  {Jackel}}]{LeCun1989}%
  \BibitemOpen
  \bibfield  {author} {\bibinfo {author} {\bibfnamefont {Y.}~\bibnamefont
  {LeCun}}, \bibinfo {author} {\bibfnamefont {B.}~\bibnamefont {Boser}},
  \bibinfo {author} {\bibfnamefont {J.~S.}\ \bibnamefont {Denker}}, \bibinfo
  {author} {\bibfnamefont {D.}~\bibnamefont {Henderson}}, \bibinfo {author}
  {\bibfnamefont {R.~E.}\ \bibnamefont {Howard}}, \bibinfo {author}
  {\bibfnamefont {W.}~\bibnamefont {Hubbard}},\ and\ \bibinfo {author}
  {\bibfnamefont {L.~D.}\ \bibnamefont {Jackel}},\ }\bibfield  {title}
  {\bibinfo {title} {{Backpropagation applied to handwritten zip code
  recognition}},\ }\href {https://doi.org/10.1162/neco.1989.1.4.541} {\bibfield
   {journal} {\bibinfo  {journal} {Neural Comput.}\ }\textbf {\bibinfo {volume}
  {1}},\ \bibinfo {pages} {541} (\bibinfo {year} {1989})}\BibitemShut {NoStop}%
\bibitem [{\citenamefont {Frandsen}\ \emph {et~al.}(2014)\citenamefont
  {Frandsen}, \citenamefont {Elesin}, \citenamefont {Frellsen}, \citenamefont
  {Mitrovic}, \citenamefont {Ding}, \citenamefont {Sigmund},\ and\
  \citenamefont {Yvind}}]{frandsen2014topology}%
  \BibitemOpen
  \bibfield  {author} {\bibinfo {author} {\bibfnamefont {L.~H.}\ \bibnamefont
  {Frandsen}}, \bibinfo {author} {\bibfnamefont {Y.}~\bibnamefont {Elesin}},
  \bibinfo {author} {\bibfnamefont {L.~F.}\ \bibnamefont {Frellsen}}, \bibinfo
  {author} {\bibfnamefont {M.}~\bibnamefont {Mitrovic}}, \bibinfo {author}
  {\bibfnamefont {Y.}~\bibnamefont {Ding}}, \bibinfo {author} {\bibfnamefont
  {O.}~\bibnamefont {Sigmund}},\ and\ \bibinfo {author} {\bibfnamefont
  {K.}~\bibnamefont {Yvind}},\ }\bibfield  {title} {\bibinfo {title} {Topology
  optimized mode conversion in a photonic crystal waveguide fabricated in
  silicon-on-insulator material},\ }\href
  {https://doi.org/10.1364/OE.22.008525} {\bibfield  {journal} {\bibinfo
  {journal} {Optics express}\ }\textbf {\bibinfo {volume} {22}},\ \bibinfo
  {pages} {8525} (\bibinfo {year} {2014})}\BibitemShut {NoStop}%
\bibitem [{\citenamefont {Lalau-Keraly}\ \emph {et~al.}(2013)\citenamefont
  {Lalau-Keraly}, \citenamefont {Bhargava}, \citenamefont {Miller},\ and\
  \citenamefont {Yablonovitch}}]{lalau2013adjoint}%
  \BibitemOpen
  \bibfield  {author} {\bibinfo {author} {\bibfnamefont {C.~M.}\ \bibnamefont
  {Lalau-Keraly}}, \bibinfo {author} {\bibfnamefont {S.}~\bibnamefont
  {Bhargava}}, \bibinfo {author} {\bibfnamefont {O.~D.}\ \bibnamefont
  {Miller}},\ and\ \bibinfo {author} {\bibfnamefont {E.}~\bibnamefont
  {Yablonovitch}},\ }\bibfield  {title} {\bibinfo {title} {Adjoint shape
  optimization applied to electromagnetic design},\ }\href
  {https://doi.org/10.1364/OE.21.021693} {\bibfield  {journal} {\bibinfo
  {journal} {Optics express}\ }\textbf {\bibinfo {volume} {21}},\ \bibinfo
  {pages} {21693} (\bibinfo {year} {2013})}\BibitemShut {NoStop}%
\bibitem [{\citenamefont {Su}\ \emph {et~al.}(2017)\citenamefont {Su},
  \citenamefont {Piggott}, \citenamefont {Sapra}, \citenamefont {Petykiewicz},\
  and\ \citenamefont {Vuckovic}}]{su2017inverse}%
  \BibitemOpen
  \bibfield  {author} {\bibinfo {author} {\bibfnamefont {L.}~\bibnamefont
  {Su}}, \bibinfo {author} {\bibfnamefont {A.~Y.}\ \bibnamefont {Piggott}},
  \bibinfo {author} {\bibfnamefont {N.~V.}\ \bibnamefont {Sapra}}, \bibinfo
  {author} {\bibfnamefont {J.}~\bibnamefont {Petykiewicz}},\ and\ \bibinfo
  {author} {\bibfnamefont {J.}~\bibnamefont {Vuckovic}},\ }\bibfield  {title}
  {\bibinfo {title} {Inverse design and demonstration of a compact on-chip
  narrowband three-channel wavelength demultiplexer},\ }\href
  {https://doi.org/10.1021/acsphotonics.7b00987} {\bibfield  {journal}
  {\bibinfo  {journal} {ACS Photonics}\ }\textbf {\bibinfo {volume} {5}},\
  \bibinfo {pages} {301} (\bibinfo {year} {2017})}\BibitemShut {NoStop}%
\bibitem [{\citenamefont {Sell}\ \emph {et~al.}(2017)\citenamefont {Sell},
  \citenamefont {Yang}, \citenamefont {Doshay}, \citenamefont {Yang},\ and\
  \citenamefont {Fan}}]{sell2017large}%
  \BibitemOpen
  \bibfield  {author} {\bibinfo {author} {\bibfnamefont {D.}~\bibnamefont
  {Sell}}, \bibinfo {author} {\bibfnamefont {J.}~\bibnamefont {Yang}}, \bibinfo
  {author} {\bibfnamefont {S.}~\bibnamefont {Doshay}}, \bibinfo {author}
  {\bibfnamefont {R.}~\bibnamefont {Yang}},\ and\ \bibinfo {author}
  {\bibfnamefont {J.~A.}\ \bibnamefont {Fan}},\ }\bibfield  {title} {\bibinfo
  {title} {Large-angle, multifunctional metagratings based on freeform
  multimode geometries},\ }\href {https://doi.org/10.1021/acs.nanolett.7b01082}
  {\bibfield  {journal} {\bibinfo  {journal} {Nano letters}\ }\textbf {\bibinfo
  {volume} {17}},\ \bibinfo {pages} {3752} (\bibinfo {year}
  {2017})}\BibitemShut {NoStop}%
\bibitem [{\citenamefont {Callewaert}\ \emph {et~al.}(2018)\citenamefont
  {Callewaert}, \citenamefont {Velev}, \citenamefont {Kumar}, \citenamefont
  {Sahakian},\ and\ \citenamefont {Aydin}}]{callewaert2018inverse}%
  \BibitemOpen
  \bibfield  {author} {\bibinfo {author} {\bibfnamefont {F.}~\bibnamefont
  {Callewaert}}, \bibinfo {author} {\bibfnamefont {V.}~\bibnamefont {Velev}},
  \bibinfo {author} {\bibfnamefont {P.}~\bibnamefont {Kumar}}, \bibinfo
  {author} {\bibfnamefont {A.}~\bibnamefont {Sahakian}},\ and\ \bibinfo
  {author} {\bibfnamefont {K.}~\bibnamefont {Aydin}},\ }\bibfield  {title}
  {\bibinfo {title} {Inverse-designed broadband all-dielectric electromagnetic
  metadevices},\ }\href {https://doi.org/10.1038/s41598-018-19796-y} {\bibfield
   {journal} {\bibinfo  {journal} {Scientific reports}\ }\textbf {\bibinfo
  {volume} {8}},\ \bibinfo {pages} {1358} (\bibinfo {year} {2018})}\BibitemShut
  {NoStop}%
\bibitem [{\citenamefont {Men}\ \emph {et~al.}(2014)\citenamefont {Men},
  \citenamefont {Lee}, \citenamefont {Freund}, \citenamefont {Peraire},\ and\
  \citenamefont {Johnson}}]{men2014robust}%
  \BibitemOpen
  \bibfield  {author} {\bibinfo {author} {\bibfnamefont {H.}~\bibnamefont
  {Men}}, \bibinfo {author} {\bibfnamefont {K.~Y.}\ \bibnamefont {Lee}},
  \bibinfo {author} {\bibfnamefont {R.~M.}\ \bibnamefont {Freund}}, \bibinfo
  {author} {\bibfnamefont {J.}~\bibnamefont {Peraire}},\ and\ \bibinfo {author}
  {\bibfnamefont {S.~G.}\ \bibnamefont {Johnson}},\ }\bibfield  {title}
  {\bibinfo {title} {Robust topology optimization of three-dimensional
  photonic-crystal band-gap structures},\ }\href
  {https://doi.org/10.1364/OE.22.022632} {\bibfield  {journal} {\bibinfo
  {journal} {Optics express}\ }\textbf {\bibinfo {volume} {22}},\ \bibinfo
  {pages} {22632} (\bibinfo {year} {2014})}\BibitemShut {NoStop}%
\bibitem [{\citenamefont {Ganapati}\ \emph {et~al.}(2014)\citenamefont
  {Ganapati}, \citenamefont {Miller},\ and\ \citenamefont
  {Yablonovitch}}]{ganapati2014light}%
  \BibitemOpen
  \bibfield  {author} {\bibinfo {author} {\bibfnamefont {V.}~\bibnamefont
  {Ganapati}}, \bibinfo {author} {\bibfnamefont {O.~D.}\ \bibnamefont
  {Miller}},\ and\ \bibinfo {author} {\bibfnamefont {E.}~\bibnamefont
  {Yablonovitch}},\ }\bibfield  {title} {\bibinfo {title} {Light trapping
  textures designed by electromagnetic optimization for subwavelength thick
  solar cells},\ }\href {https://doi.org/10.1109/JPHOTOV.2013.2280340}
  {\bibfield  {journal} {\bibinfo  {journal} {IEEE Journal of Photovoltaics}\
  }\textbf {\bibinfo {volume} {4}},\ \bibinfo {pages} {175} (\bibinfo {year}
  {2014})}\BibitemShut {NoStop}%
\bibitem [{\citenamefont {Shen}\ \emph {et~al.}(2014)\citenamefont {Shen},
  \citenamefont {Wang}, \citenamefont {Polson},\ and\ \citenamefont
  {Menon}}]{shen2014ultra}%
  \BibitemOpen
  \bibfield  {author} {\bibinfo {author} {\bibfnamefont {B.}~\bibnamefont
  {Shen}}, \bibinfo {author} {\bibfnamefont {P.}~\bibnamefont {Wang}}, \bibinfo
  {author} {\bibfnamefont {R.}~\bibnamefont {Polson}},\ and\ \bibinfo {author}
  {\bibfnamefont {R.}~\bibnamefont {Menon}},\ }\bibfield  {title} {\bibinfo
  {title} {Ultra-high-efficiency metamaterial polarizer},\ }\href
  {https://doi.org/10.1364/OPTICA.1.000356} {\bibfield  {journal} {\bibinfo
  {journal} {Optica}\ }\textbf {\bibinfo {volume} {1}},\ \bibinfo {pages} {356}
  (\bibinfo {year} {2014})}\BibitemShut {NoStop}%
\bibitem [{\citenamefont {Lin}\ \emph {et~al.}(2019)\citenamefont {Lin},
  \citenamefont {Liu}, \citenamefont {Pestourie},\ and\ \citenamefont
  {Johnson}}]{lin2019topology}%
  \BibitemOpen
  \bibfield  {author} {\bibinfo {author} {\bibfnamefont {Z.}~\bibnamefont
  {Lin}}, \bibinfo {author} {\bibfnamefont {V.}~\bibnamefont {Liu}}, \bibinfo
  {author} {\bibfnamefont {R.}~\bibnamefont {Pestourie}},\ and\ \bibinfo
  {author} {\bibfnamefont {S.~G.}\ \bibnamefont {Johnson}},\ }\bibfield
  {title} {\bibinfo {title} {Topology optimization of freeform large-area
  metasurfaces},\ }\bibfield  {journal} {\bibinfo  {journal} {arXiv preprint
  arXiv:1902.03179}\ }\href {https://doi.org/arXiv:1902.03179}
  {arXiv:1902.03179} (\bibinfo {year} {2019})\BibitemShut {NoStop}%
\bibitem [{\citenamefont {Liu}\ and\ \citenamefont {Barnett}(2016)}]{Liu2016}%
  \BibitemOpen
  \bibfield  {author} {\bibinfo {author} {\bibfnamefont {Y.}~\bibnamefont
  {Liu}}\ and\ \bibinfo {author} {\bibfnamefont {A.~H.}\ \bibnamefont
  {Barnett}},\ }\bibfield  {title} {\bibinfo {title} {{Efficient numerical
  solution of acoustic scattering from doubly-periodic arrays of axisymmetric
  objects}},\ }\href {https://doi.org/10.1016/j.jcp.2016.08.011} {\bibfield
  {journal} {\bibinfo  {journal} {J. Comput. Phys.}\ }\textbf {\bibinfo
  {volume} {324}},\ \bibinfo {pages} {226} (\bibinfo {year}
  {2016})}\BibitemShut {NoStop}%
\bibitem [{\citenamefont {Bruno}\ and\ \citenamefont {Maas}(2019)}]{Bruno2019}%
  \BibitemOpen
  \bibfield  {author} {\bibinfo {author} {\bibfnamefont {O.~P.}\ \bibnamefont
  {Bruno}}\ and\ \bibinfo {author} {\bibfnamefont {M.}~\bibnamefont {Maas}},\
  }\bibfield  {title} {\bibinfo {title} {{Shifted equivalent sources and FFT
  acceleration for periodic scattering problems, including Wood anomalies}},\
  }\href {https://doi.org/10.1016/j.jcp.2018.10.044} {\bibfield  {journal}
  {\bibinfo  {journal} {J. Comput. Phys.}\ }\textbf {\bibinfo {volume} {378}},\
  \bibinfo {pages} {548} (\bibinfo {year} {2019})}\BibitemShut {NoStop}%
\bibitem [{\citenamefont {Trefethen}\ and\ \citenamefont {{Bau
  III}}(1997)}]{Trefethen1997}%
  \BibitemOpen
  \bibfield  {author} {\bibinfo {author} {\bibfnamefont {L.~N.}\ \bibnamefont
  {Trefethen}}\ and\ \bibinfo {author} {\bibfnamefont {D.}~\bibnamefont {{Bau
  III}}},\ }\href@noop {} {\emph {\bibinfo {title} {{Numerical linear
  algebra}}}}\ (\bibinfo  {publisher} {SIAM},\ \bibinfo {year}
  {1997})\BibitemShut {NoStop}%
\bibitem [{\citenamefont {Swanson}(1989)}]{swanson1989binary}%
  \BibitemOpen
  \bibfield  {author} {\bibinfo {author} {\bibfnamefont {G.}~\bibnamefont
  {Swanson}},\ }\bibfield  {title} {\bibinfo {title} {Binary optics technology:
  the theory and design of multilevel diffractive optical elements},\
  }\href@noop {} {\bibfield  {journal} {\bibinfo  {journal} {MIT Lincoln
  Laboratory Technical Report 854}\ } (\bibinfo {year} {1989})}\BibitemShut
  {NoStop}%
\bibitem [{\citenamefont {Kamali}\ \emph {et~al.}(2018)\citenamefont {Kamali},
  \citenamefont {Arbabi}, \citenamefont {Arbabi},\ and\ \citenamefont
  {Faraon}}]{kamali2018review}%
  \BibitemOpen
  \bibfield  {author} {\bibinfo {author} {\bibfnamefont {S.~M.}\ \bibnamefont
  {Kamali}}, \bibinfo {author} {\bibfnamefont {E.}~\bibnamefont {Arbabi}},
  \bibinfo {author} {\bibfnamefont {A.}~\bibnamefont {Arbabi}},\ and\ \bibinfo
  {author} {\bibfnamefont {A.}~\bibnamefont {Faraon}},\ }\bibfield  {title}
  {\bibinfo {title} {A review of dielectric optical metasurfaces for wavefront
  control},\ }\href {https://doi.org/10.1515/nanoph-2017-0129} {\bibfield
  {journal} {\bibinfo  {journal} {Nanophotonics}\ }\textbf {\bibinfo {volume}
  {7}},\ \bibinfo {pages} {1041} (\bibinfo {year} {2018})}\BibitemShut
  {NoStop}%
\bibitem [{\citenamefont {Palik}(1998)}]{palik1998handbook}%
  \BibitemOpen
  \bibfield  {author} {\bibinfo {author} {\bibfnamefont {E.~D.}\ \bibnamefont
  {Palik}},\ }\href@noop {} {\emph {\bibinfo {title} {Handbook of optical
  constants of solids}}},\ Vol.~\bibinfo {volume} {3}\ (\bibinfo  {publisher}
  {Academic press},\ \bibinfo {year} {1998})\BibitemShut {NoStop}%
\bibitem [{\citenamefont {Kong}(1975)}]{kong1975theory}%
  \BibitemOpen
  \bibfield  {author} {\bibinfo {author} {\bibfnamefont {J.~A.}\ \bibnamefont
  {Kong}},\ }\bibfield  {title} {\bibinfo {title} {Theory of electromagnetic
  waves},\ }\href@noop {} {\bibfield  {journal} {\bibinfo  {journal} {New York,
  Wiley-Interscience, 1975. 348 p.}\ } (\bibinfo {year} {1975})}\BibitemShut
  {NoStop}%
\bibitem [{\citenamefont {Johnson}\ \emph {et~al.}(2002)\citenamefont
  {Johnson}, \citenamefont {Ibanescu}, \citenamefont {Skorobogatiy},
  \citenamefont {Weisberg}, \citenamefont {Joannopoulos},\ and\ \citenamefont
  {Fink}}]{johnson2002perturbation}%
  \BibitemOpen
  \bibfield  {author} {\bibinfo {author} {\bibfnamefont {S.~G.}\ \bibnamefont
  {Johnson}}, \bibinfo {author} {\bibfnamefont {M.}~\bibnamefont {Ibanescu}},
  \bibinfo {author} {\bibfnamefont {M.}~\bibnamefont {Skorobogatiy}}, \bibinfo
  {author} {\bibfnamefont {O.}~\bibnamefont {Weisberg}}, \bibinfo {author}
  {\bibfnamefont {J.}~\bibnamefont {Joannopoulos}},\ and\ \bibinfo {author}
  {\bibfnamefont {Y.}~\bibnamefont {Fink}},\ }\bibfield  {title} {\bibinfo
  {title} {Perturbation theory for {M}axwell's equations with shifting material
  boundaries},\ }\href {https://doi.org/10.1103/PhysRevE.65.066611} {\bibfield
  {journal} {\bibinfo  {journal} {Physical review E}\ }\textbf {\bibinfo
  {volume} {65}},\ \bibinfo {pages} {066611} (\bibinfo {year}
  {2002})}\BibitemShut {NoStop}%
\bibitem [{\citenamefont {Oskooi}\ \emph {et~al.}(2010)\citenamefont {Oskooi},
  \citenamefont {Roundy}, \citenamefont {Ibanescu}, \citenamefont {Bermel},
  \citenamefont {Joannopoulos},\ and\ \citenamefont
  {Johnson}}]{oskooi2010meep}%
  \BibitemOpen
  \bibfield  {author} {\bibinfo {author} {\bibfnamefont {A.~F.}\ \bibnamefont
  {Oskooi}}, \bibinfo {author} {\bibfnamefont {D.}~\bibnamefont {Roundy}},
  \bibinfo {author} {\bibfnamefont {M.}~\bibnamefont {Ibanescu}}, \bibinfo
  {author} {\bibfnamefont {P.}~\bibnamefont {Bermel}}, \bibinfo {author}
  {\bibfnamefont {J.~D.}\ \bibnamefont {Joannopoulos}},\ and\ \bibinfo {author}
  {\bibfnamefont {S.~G.}\ \bibnamefont {Johnson}},\ }\bibfield  {title}
  {\bibinfo {title} {Meep: A flexible free-software package for electromagnetic
  simulations by the fdtd method},\ }\href
  {https://doi.org/10.1016/j.cpc.2009.11.008} {\bibfield  {journal} {\bibinfo
  {journal} {Computer Physics Communications}\ }\textbf {\bibinfo {volume}
  {181}},\ \bibinfo {pages} {687} (\bibinfo {year} {2010})}\BibitemShut
  {NoStop}%
\bibitem [{\citenamefont {Taflove}\ \emph {et~al.}(2013)\citenamefont
  {Taflove}, \citenamefont {Oskooi},\ and\ \citenamefont
  {Johnson}}]{taflove2013advances}%
  \BibitemOpen
  \bibfield  {author} {\bibinfo {author} {\bibfnamefont {A.}~\bibnamefont
  {Taflove}}, \bibinfo {author} {\bibfnamefont {A.}~\bibnamefont {Oskooi}},\
  and\ \bibinfo {author} {\bibfnamefont {S.~G.}\ \bibnamefont {Johnson}},\
  }\href@noop {} {\emph {\bibinfo {title} {Advances in FDTD computational
  electrodynamics: photonics and nanotechnology}}}\ (\bibinfo  {publisher}
  {Artech house},\ \bibinfo {year} {2013})\BibitemShut {NoStop}%
\bibitem [{\citenamefont {Boyd}\ and\ \citenamefont
  {Vandenberghe}(2004)}]{boyd2004convex}%
  \BibitemOpen
  \bibfield  {author} {\bibinfo {author} {\bibfnamefont {S.}~\bibnamefont
  {Boyd}}\ and\ \bibinfo {author} {\bibfnamefont {L.}~\bibnamefont
  {Vandenberghe}},\ }\href@noop {} {\emph {\bibinfo {title} {Convex
  optimization}}}\ (\bibinfo  {publisher} {Cambridge university press},\
  \bibinfo {year} {2004})\BibitemShut {NoStop}%
\bibitem [{\citenamefont {Ferreira}\ and\ \citenamefont
  {Kempf}(2006)}]{Ferreira2006}%
  \BibitemOpen
  \bibfield  {author} {\bibinfo {author} {\bibfnamefont {P.~J.}\ \bibnamefont
  {Ferreira}}\ and\ \bibinfo {author} {\bibfnamefont {A.}~\bibnamefont
  {Kempf}},\ }\bibfield  {title} {\bibinfo {title} {{Superoscillations: Faster
  than the Nyquist rate}},\ }\href {https://doi.org/10.1109/TSP.2006.877642}
  {\bibfield  {journal} {\bibinfo  {journal} {IEEE Transactions on Signal
  Processing}\ }\textbf {\bibinfo {volume} {54}},\ \bibinfo {pages} {3732}
  (\bibinfo {year} {2006})}\BibitemShut {NoStop}%
\bibitem [{\citenamefont {Shim}\ \emph {et~al.}(2019)\citenamefont {Shim},
  \citenamefont {Chung},\ and\ \citenamefont {Miller}}]{Shim2019}%
  \BibitemOpen
  \bibfield  {author} {\bibinfo {author} {\bibfnamefont {H.}~\bibnamefont
  {Shim}}, \bibinfo {author} {\bibfnamefont {H.}~\bibnamefont {Chung}},\ and\
  \bibinfo {author} {\bibfnamefont {O.~D.}\ \bibnamefont {Miller}},\
  }\href@noop {} {\bibfield  {journal} {\bibinfo  {journal} {To be published}\
  } (\bibinfo {year} {2019})}\BibitemShut {NoStop}%
\bibitem [{\citenamefont {Wang}\ \emph {et~al.}(2016)\citenamefont {Wang},
  \citenamefont {Mohammad},\ and\ \citenamefont {Menon}}]{wang2016chromatic}%
  \BibitemOpen
  \bibfield  {author} {\bibinfo {author} {\bibfnamefont {P.}~\bibnamefont
  {Wang}}, \bibinfo {author} {\bibfnamefont {N.}~\bibnamefont {Mohammad}},\
  and\ \bibinfo {author} {\bibfnamefont {R.}~\bibnamefont {Menon}},\ }\bibfield
   {title} {\bibinfo {title} {Chromatic-aberration-corrected diffractive lenses
  for ultra-broadband focusing},\ }\href {https://doi.org/10.1038/srep21545}
  {\bibfield  {journal} {\bibinfo  {journal} {Scientific reports}\ }\textbf
  {\bibinfo {volume} {6}},\ \bibinfo {pages} {21545} (\bibinfo {year}
  {2016})}\BibitemShut {NoStop}%
\end{thebibliography}
%
\end{document}


\title{Supplementary Materials: High-NA Achromatic Metalenses by Inverse Design}

\author{Haejun Chung}
\affiliation{Department of Applied Physics and Energy Sciences Institute, Yale University, New Haven, Connecticut 06511, USA}
\author{Owen D. Miller}
\affiliation{Department of Applied Physics and Energy Sciences Institute, Yale University, New Haven, Connecticut 06511, USA}

\date{\today}

\maketitle
\tableofcontents

\section{Material-dispersion simulation verification}
\begin{figure}[ht]
\centering
\includegraphics[width=0.5\linewidth]{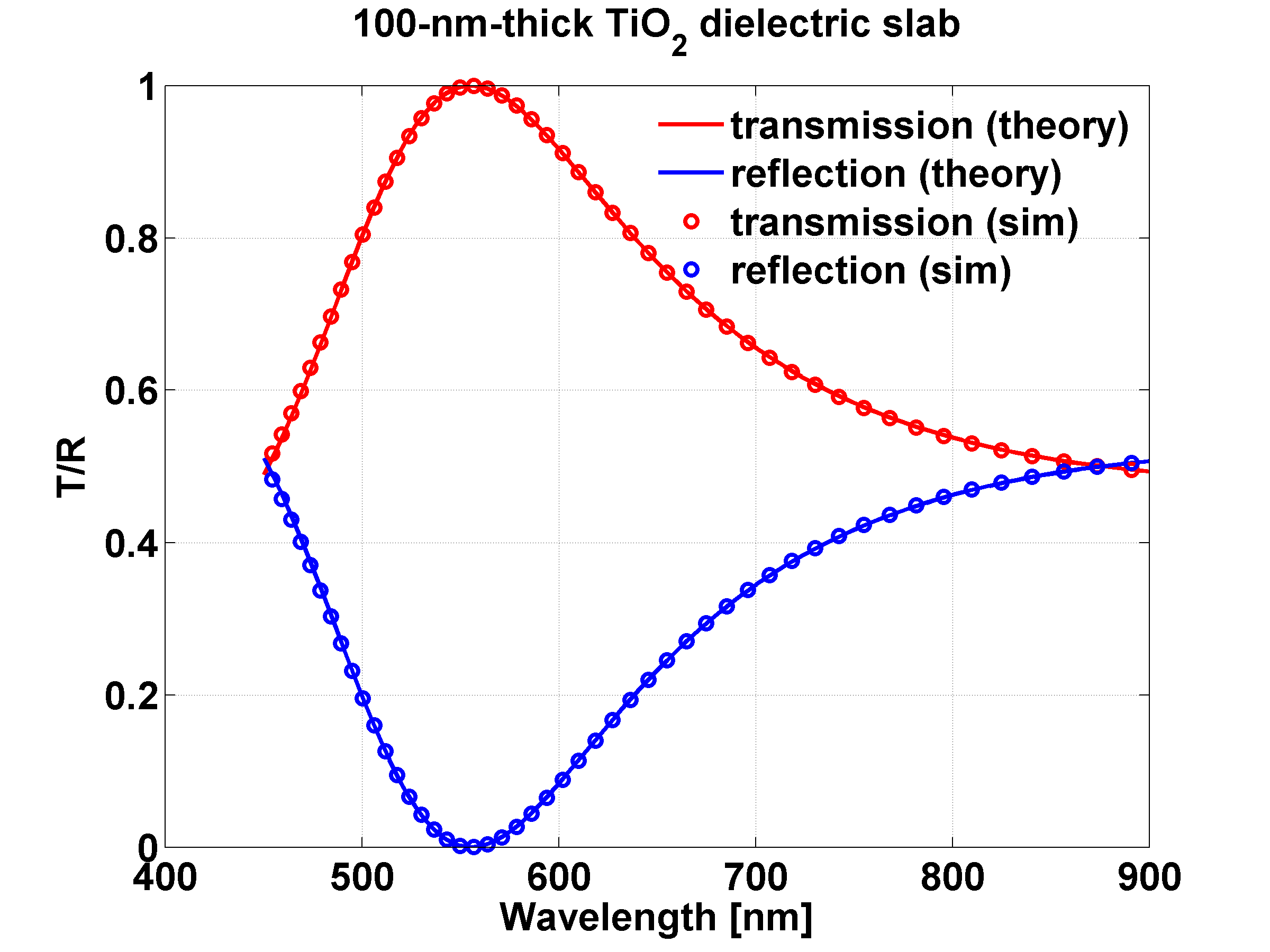}
\caption{Simulated and calculated reflection and transmission of 100-nm-thick TiO$_2$ dielectric slab. They match very well over the visible spectrum.}
\label{analytic_absorption}
\end{figure}

In this work, Finite Difference Time Domain (FDTD) method is used to solve a full Maxwell equation~\cite{oskooi2010meep,taflove2013advances}. FDTD can cover a wide range of frequencies with a single time domain simulation. However, it requires time-domain modeling of material dispersion. We model the dispersion of TiO$_2$ material with Drude model and validate the material modeling with a theoretical calculation of Fabry-Perot oscillations in a 1D dielectric slab~\cite{orfanidis2002electromagnetic} as shown in Fig.~\ref{analytic_absorption}.

\newpage
\section{Achromatic metalenses with a glass substrate}
\begin{figure*}[h]
\centering
\includegraphics[width=0.8\linewidth]{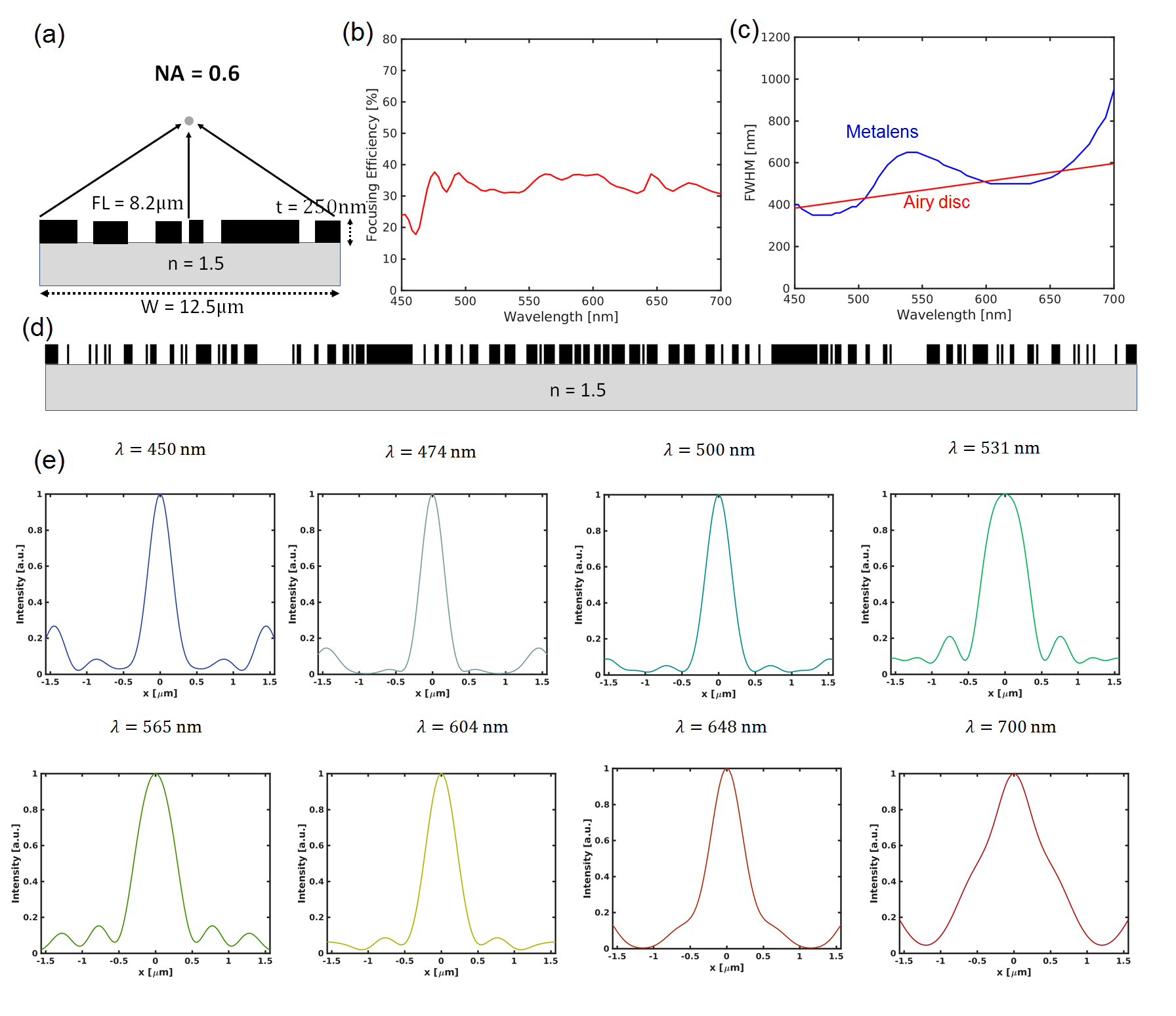}
\caption{Achromatic metalens for the visible wavelength (450 -- 700 nm) with substrate (n = 1.5) (a) Schematic illustration of metalens dimension. (b) Focusing efficiency of the optimized metalens. The average efficiency is 32 \% over the visible spectrum. The efficiency was calculated by the power within the first minimum point divided by the incidence power. (c) Full Width Half Maximum (FWHM) of the optimized metalens and the corresponding airy disk. (d) Optimized TiO$_2$ based metalens structure. The width and thickness of the metalens are 12.5 $\mu$m and 250 nm, respectively. 25 nm grid spacing was used. The maximum aspect ratio is 10. (e) Normalized intensity ($|$E$|$$^2$) profile.}
\label{supple1}
\end{figure*}

Metalenses are generally fabricated on transparent substrates. Here, we demonstrate achromatic metalenses with SiO$_2$ substrate (n = 1.5). The incidence power is calculated within the glass substrate. As shown in Fig.~\ref{supple1} (a), the optimized device has 250 nm thickness, 12.5 $\mu$m width, and 8.2 $\mu$m focal length, which corresponds to NA = 0.6. As shown in Fig.~\ref{supple1} (b), a calculated average focusing efficiency is 32 \% over visible wavelengths
(450 – 700 nm) while the highest efficiency is 39 \%. The focusing efficiency is defined as the ratio of the E-field intensity within the first minimum point and the incidence
power~\cite{chen2018broadband}. As shown in Fig.~\ref{supple1}(c),
the calculated full-width half-maximum (FWHM) over visible
wavelength is very close to the diffraction limited FWHM except for the longer wavelength.
Figure.~\ref{supple1}(d) shows the
optimized structure which is feasible to existing lithography
techniques~\cite{owa2004advantage,khorasaninejad2016polarization}. Figure.~\ref{supple1}(e) shows E-field intensity profiles at the focal plane. Except for the 700nm wavelength, they are very close to diffraction limited focal spots.

\newpage
\section{Least-squares optimal coefficients}
In the main text, we solve for optimal unit-cell coefficients to achieve a desired field distribution. In the setup of the problem, there is a given field $\Ev(\xv)$ that can be decomposed into an orthonormal basis of functions $\Ev_i(\xv)$ and coefficients $c_i$, and a target field $\Ev_{\rm tar}(\xv)$ that one would like $\Ev(\xv)$ to most closely approximately, i.e., there is a functional $F$ of the coefficients $c_i$ to be minimized:
\begin{align}
    \min_{c_i} F(c_i) = \min_{c_i} \left\| \Ev_{\rm tar} - \Ev(\xv) \right\|_2^2 = \min_{c_i} \left\| \Ev_{\rm tar} - \sum_i c_i \Ev_i(\xv) \right\|_2^2.
\end{align}
In the main text we stated that the optimal coefficients are given by
\begin{align}
    c_i = \int \cc{\Ev}_i \cdot \Ev_{\rm tar}.
    \label{copt}
\end{align}
This statement is a well-known result for linear least squares problems. Here, we provide a short proof in this context.

For any given $i$, we can compute the value of $c_i$ for which $\partial F / \partial c_i$ equals 0. In fact, we must carefully consider both the real and imaginary parts of $c_i$; instead, we can use the complex CR calculus~\cite{KreutzDelgado2009}, whereby $c_i$ and its conjugate $\cc{c_i}$ are formally considered independent variables, and the derivatives of $F$ with respect to each must be zero. Since $F$ is real-valued the two derivatives contain redundant information, and only one must be set to zero; for simplicity in the ultimately derivation, we choose the derivative with respect to $\cc{c_i}$:
\begin{align}
    \frac{\partial F}{\partial \cc{c_i}} &=  - \int \cc{\Ev}_i \cdot \left( \Ev_{\rm tar} - \sum_j c_j \Ev_j \right) \nonumber \\
                                         &= -\int \cc{\Ev}_i \cdot \Ev_{\rm tar} + c_i,
\end{align}
where in the latter expression we have used the orthonormality of the basis functions, $\int \cc{\Ev}_j \cdot \Ev_i = \delta_{ij}$. Thus the condition for zero first derivative requires Eq.~\ref{copt} to hold; as the only extremum it must also be the minimum, thus proving that it is the solution to the minimization problem.

\newpage
\section{Summary of optimized metalenses for various numerical apertures using translation invariant geometry}
\begin{figure*}[ht]
\centering
\includegraphics[width=\linewidth]{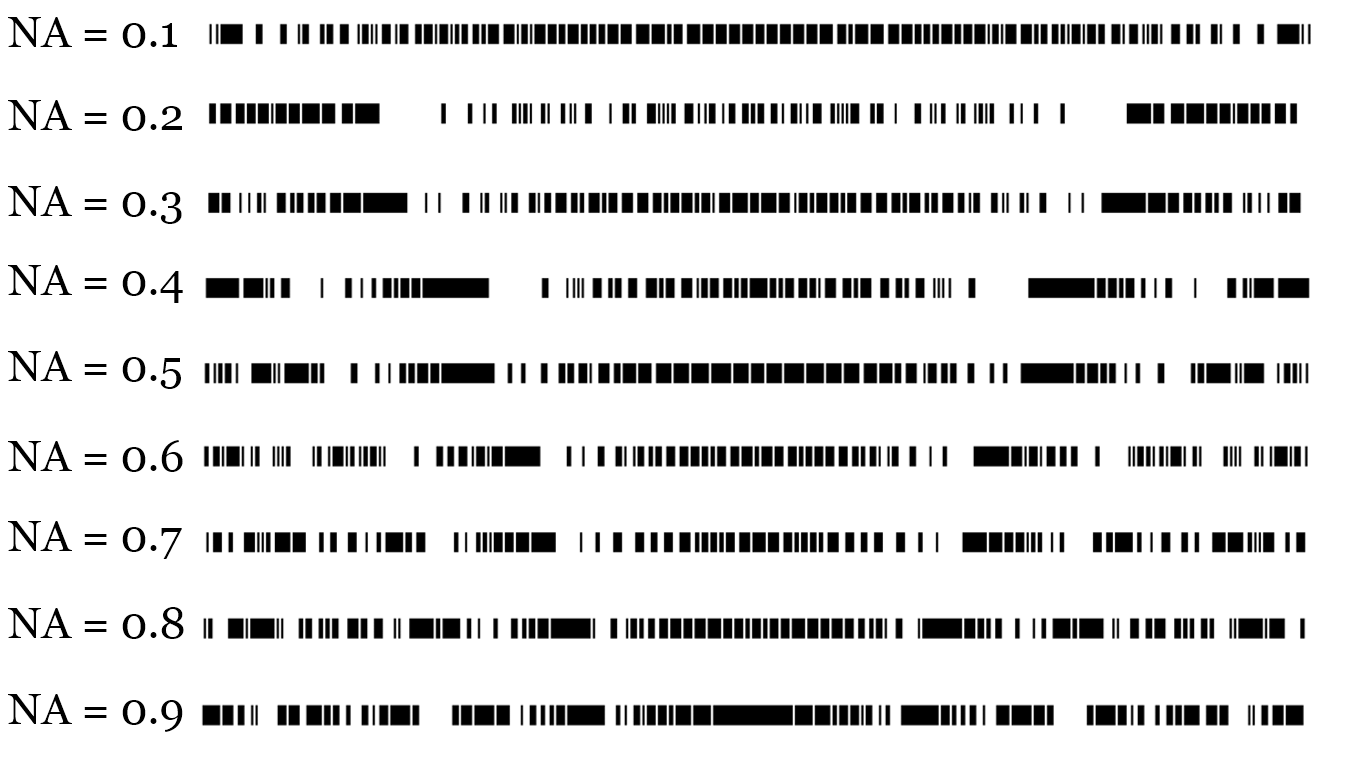}
\caption{Optimized metalenses for various NA (0.1 -- 0.9) using translation invariant geometry. The thickness and width are 250 nm and 12.5 $\mu$m, respectively. The minimum resolution is 25 nm with the maximum aspect ratio of 10.}
\label{supple2}
\end{figure*}

\begin{figure*}[ht]
\centering
\includegraphics[width=\linewidth]{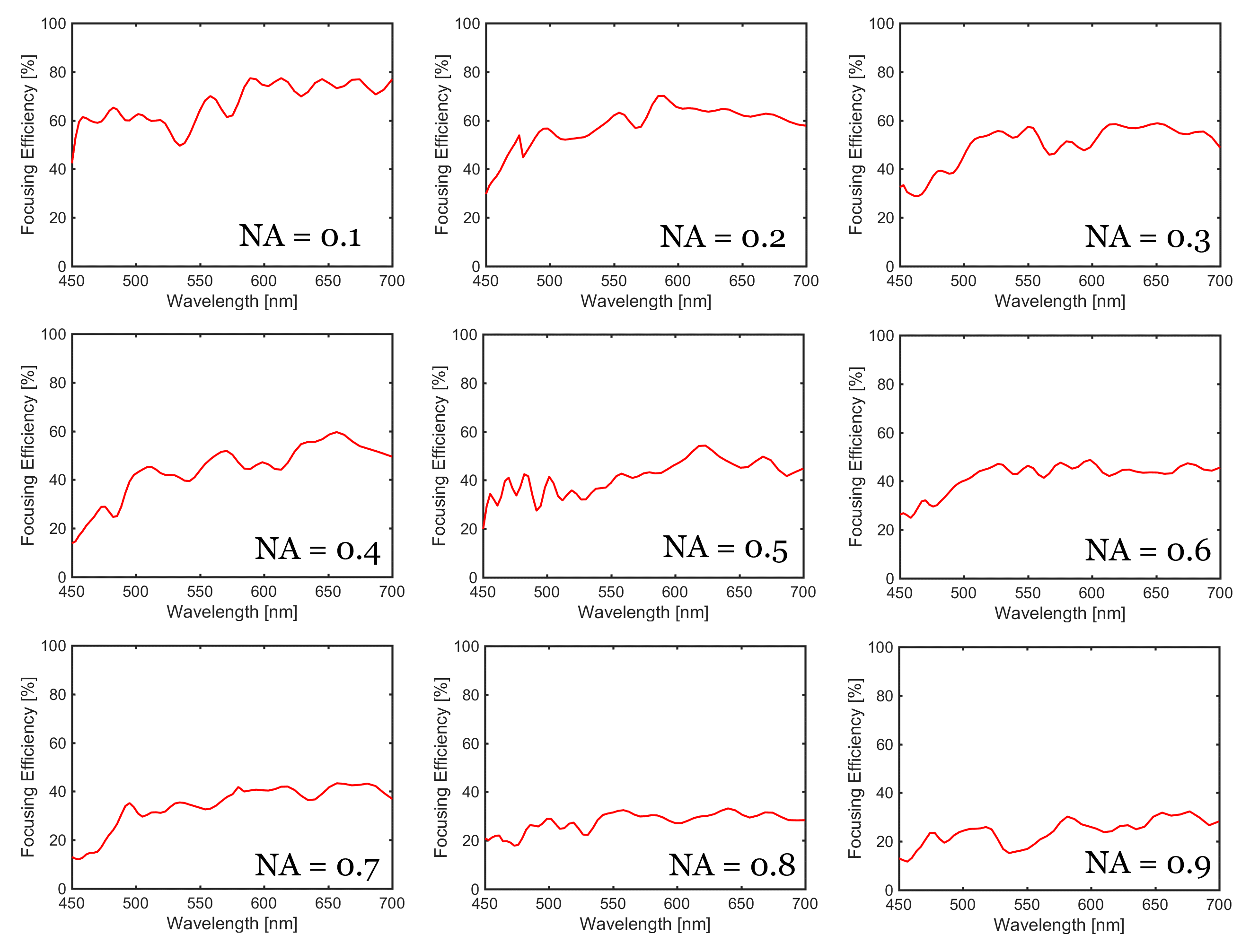}
\caption{Achieved efficiency over 450 -- 700 nm wavelength range for various NA (0.1 -- 0.9) using translation invariant geometry. The blue curves are focusing efficiencies (as defined in the main text) while the red curves are efficiency within diffraction limit, which is calculated via ratio between power within diffraction limit and incidence power.}
\label{supple3}
\end{figure*}

We demonstrated low to high NA metalenses in the main text. Here, we attach detailed optimized structures and efficiencies at the visible wavelength. As shown in Fig.~\ref{supple2}, the optimized structures, in the macroscopic view, have a gradually varying TiO$_2$ filling ratio over device radial direction. The speed of this variation increases over increasing NA. Figure~\ref{supple3} shows achieved efficiencies for various NA (0.1 – 0.9) using translation invariant geometry. The detailed geometric parameters are attached at the end of this supplementary material.

\newpage

\newpage

\section{Summary of optimized metalenses for various numerical apertures using freeform geometry}

\begin{figure*}[ht]
\centering
\includegraphics[width=\linewidth]{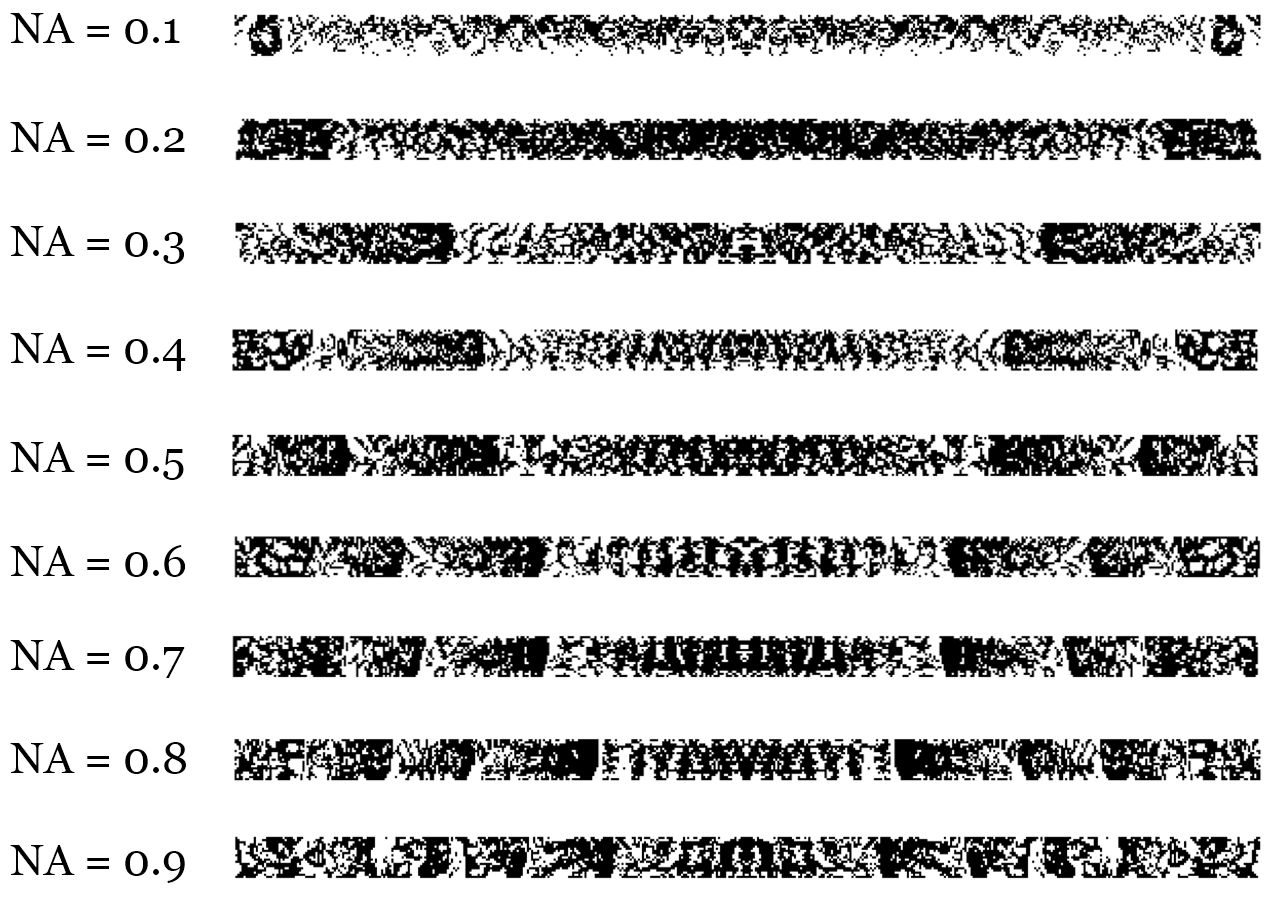}
\caption{Optimized metalenses for various NA (0.1 -- 0.9) using freeform geometry. The thickness and width are 500 nm and 12.5 $\mu$m, respectively. The minimum resolution is 25 nm.}
\label{supple4}
\end{figure*}

\begin{figure*}[ht]
\centering
\includegraphics[width=\linewidth]{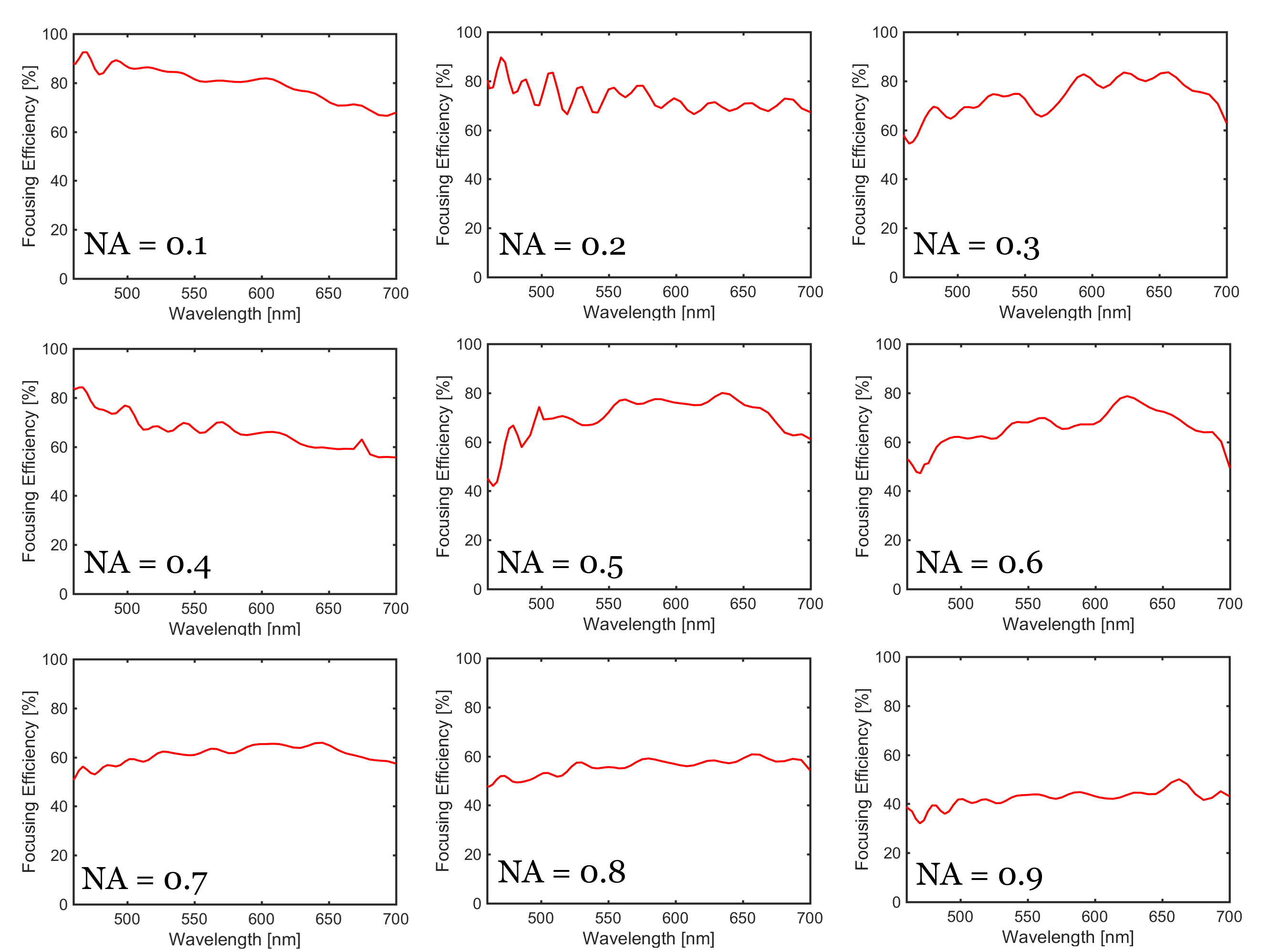}
\caption{Achieved efficiency over 450 -- 700 nm wavelength range for various NA (0.1 -- 0.9) using freeform geometry. The blue curves are focusing efficiencies (as defined in the main text) while the red curves are efficiency within diffraction limit, which is calculated via ratio between power within diffraction limit and incidence power.}
\label{supple5}
\end{figure*}

The numerical aperture (NA) versus efficiencies plot is shown in the main text. The optimized geometries and efficiencies data are shown in Figs.~\ref{supple4},~\ref{supple5}. It might be hard to figure out any physical insight from the optimized geometries. Again, in the macroscopic view, they seem to have a gradually varying TiO$_2$ filling ratio over device radial direction. The detailed geometric parameters are attached in the separate supplementary text files.

\newpage

\newpage

\section{Analysis of larger device size}
\begin{figure}[ht]
\centering
\includegraphics[width=0.5\linewidth]{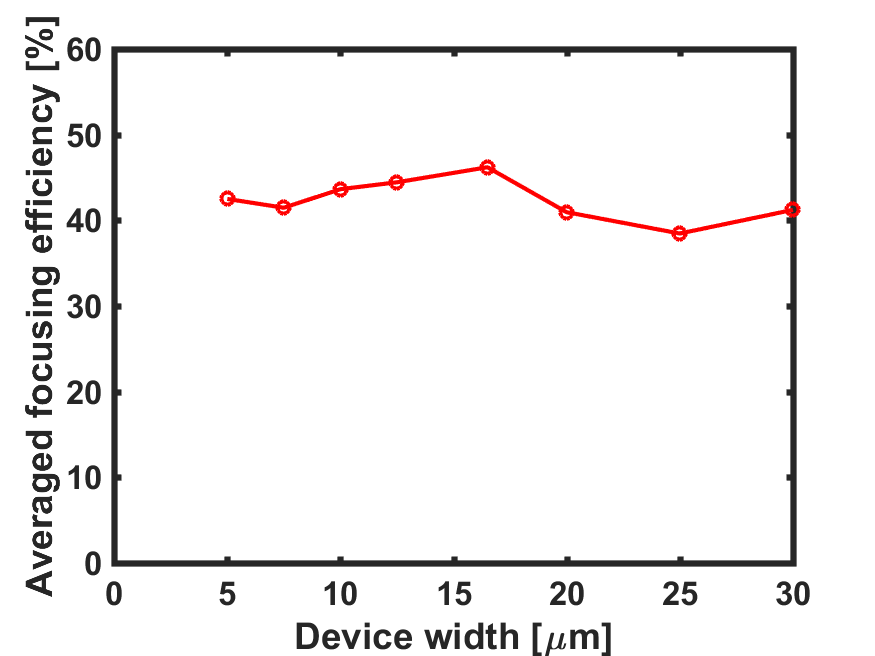}
\caption{Average focusing efficiencies for different device sizes.}
\label{supple_device_size}
\end{figure}

Large area design is a particular problem in the metalens community~\cite{kamali2018review}. To address this issue, we optimize metalenses with different device widths from 10 $\lambda$ to 60 $\lambda$. NA is fixed to 0.3 and then the device width is adjusted to see the focusing efficiency variation. The optimized structure is translational invariant geometry with the 250-nm-thick TiO$_2$ material. As shown in Fig.~\ref{supple_device_size}, the average efficiencies are close to 40 \% for different device sizes. There is no sign of significant efficiency drop for larger devices. Note that it was tested up to the 60 $\lambda$ long device where our computational capacity can handle.

\newpage

\section{Frequency sensitivity study}
\begin{figure}[ht]
\centering
\includegraphics[width=0.5\linewidth]{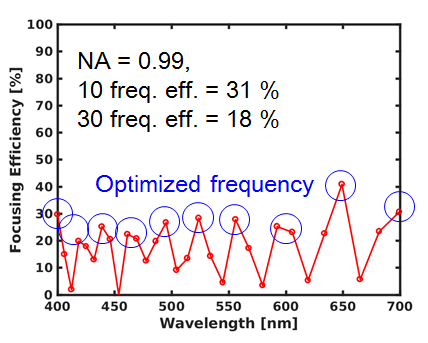}
\caption{Focusing efficiency versus wavelength. The optimization was initially performed over 10 discrete wavelengths (blue circled data), then the performance of metalens was plotted over 30 wavelengths (red curve) to check the broadband operation where the efficiency dropped from 31 \% (10 wavelengths calculation) to 18 \% (30 wavelengths calculation).}
\label{supple_freq_sensitivity}
\end{figure}

Early metalens works have demonstrated discrete operating wavelengths~\cite{aieta2015multiwavelength,khorasaninejad2015achromatic,khorasaninejad2016polarization}. Then, in order to achieve a pure achromatic operation, all of phase profile, group delay, and group delay dispersion are modeled by the unit cell approach~\cite{chen2018broadband}. Also, densely selected wavelengths~\cite{wang2018broadband,avayu2017composite,mohammad2018broadband} could also realize a continuous achromatic metalenses. The former should work well for a certain bandwidth centered at the selected wavelength. However, many studies using the latter method do not check the performance over continuous bandwidth. Here, we optimize an achromatic metalens with 10 discrete wavelengths ranging from 450 nm to 700 nm for NA = 0.99, which corresponds to $\Delta\lambda =$ 35 nm. As shown in Fig.~\ref{supple_freq_sensitivity}, the selected ten wavelengths show 31 \% average efficiency while the average efficiency calculated with finer $\Delta\lambda$ is only about 18 \%. This implies that many existing achromatic metalenses which were evaluated at well separated wavelengths may only work at those wavelengths.

\newpage

\section{Metalens design in 3D}
\begin{figure}[ht]
\centering
\includegraphics[width=0.8\linewidth]{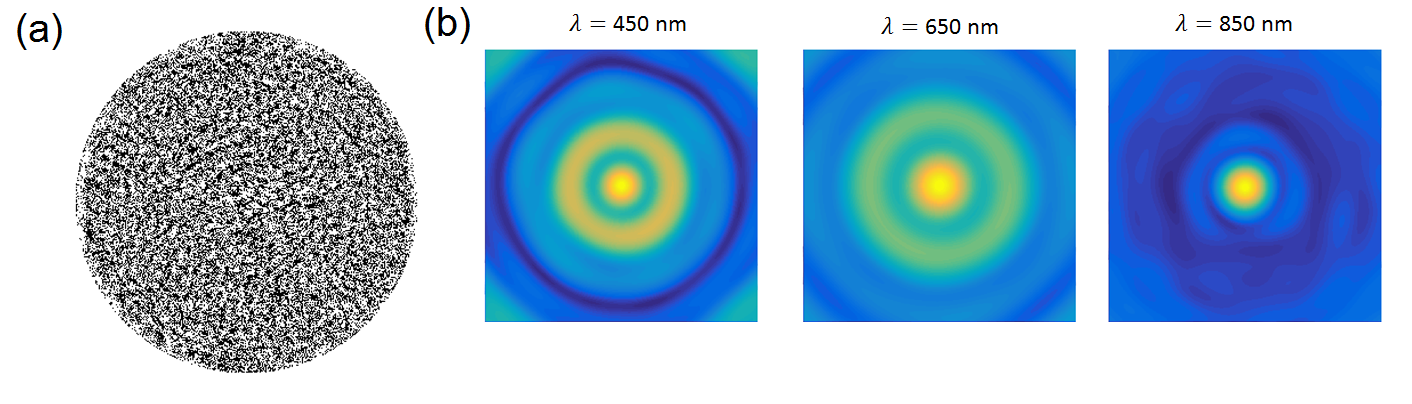}
\caption{Optimized 3D metalens. Thickness of device is 0.25 um and lens diameter is 7.5 $\mu$m and NA is 0.25 (a) top view of the 3D metalens. (b) E-field intensity at the focal plane.}
\label{supple_3D}
\end{figure}

In the main text, we demonstrated 2D achromatic metalenses due to the limit of our computational resource. Here, we present a relatively small 3D achromatic Metalens to validate our methodology in 3D. The diameter of the lens is 7.5 $\mu$m and thickness is 250 nm. The focal length is 15.0 $\mu$m which corresponds to NA = 0.25. Figure.~\ref{supple_3D} (a) shows the optimized 3D achromatic metalens in top view. The black region indicates TiO$_2$ and the white region indicates air. The focal length remains nearly constant at three target frequencies (450, 650, 850 nm). The electric-field intensity profiles at the focal plane (z = 15.0 $\mu$m) is shown in Fig.~\ref{supple_3D}(b). Note that this is a preliminary result of designing 3D achromatic metalens. To further improve this result, it may require (1) smaller feature sizes (2) a better angular resolution in order decomposition (3) a good guess on initial geometry parameters. The two formers can be resolved with a greater computing resource while the latter may need theoretical study~\cite{angeris2019computational}. 

\begin{center}
\begin{table}
\caption{Constant-z geometry parameters of every grid point (25-nm-long) of the nine metasurfaces in Fig. 1(c). The grid number starts from the left side of the metasurface. Value of one indicates TiO$_2$ and zero indicates air.}
\begin{tabular}{ |c|c|c|c|c|c|c|c|c|c| } 
\hline
Grid Number & NA=0.1 & NA=0.2 & NA=0.3 & NA=0.4 & NA=0.5 & NA=0.6 & NA=0.7 & NA=0.8 & NA=0.9 \\
\hline
1 &  0 &  0 &  0 &  1 &  1 &  1 &  1 &  1 &  0 \\ 
2 &  1 &  0 &  0 &  1 &  1 &  1 &  0 &  0 &  1 \\ 
3 &  0 &  0 &  0 &  1 &  0 &  0 &  0 &  1 &  1 \\ 
4 &  0 &  0 &  1 &  1 &  0 &  0 &  1 &  1 &  1 \\ 
5 &  1 &  1 &  1 &  1 &  1 &  1 &  1 &  0 &  1 \\ 
6 &  0 &  1 &  1 &  1 &  0 &  1 &  1 &  0 &  1 \\ 
7 &  1 &  1 &  1 &  1 &  1 &  1 &  1 &  0 &  1 \\ 
8 &  1 &  0 &  1 &  1 &  1 &  0 &  0 &  0 &  1 \\ 
9 &  1 &  0 &  0 &  1 &  0 &  1 &  0 &  0 &  1 \\ 
10 &  1 &  1 &  1 &  1 &  1 &  0 &  0 &  0 &  0 \\ 
11 &  1 &  1 &  1 &  1 &  1 &  1 &  1 &  0 &  1 \\ 
12 &  1 &  1 &  1 &  1 &  1 &  1 &  1 &  1 &  1 \\ 
13 &  1 &  1 &  1 &  1 &  0 &  1 &  0 &  1 &  1 \\ 
14 &  1 &  1 &  0 &  1 &  0 &  1 &  0 &  1 &  1 \\ 
15 &  1 &  0 &  0 &  1 &  1 &  1 &  0 &  1 &  1 \\ 
16 &  1 &  0 &  0 &  0 &  0 &  1 &  0 &  1 &  0 \\ 
17 &  0 &  1 &  0 &  0 &  0 &  0 &  0 &  1 &  0 \\ 
18 &  0 &  1 &  1 &  1 &  0 &  1 &  1 &  1 &  1 \\ 
19 &  0 &  1 &  0 &  1 &  0 &  0 &  1 &  0 &  1 \\ 
20 &  0 &  1 &  0 &  1 &  0 &  0 &  1 &  1 &  1 \\ 
21 &  0 &  0 &  0 &  1 &  0 &  0 &  1 &  0 &  0 \\ 
22 &  0 &  1 &  1 &  1 &  1 &  1 &  1 &  1 &  0 \\ 
23 &  1 &  1 &  0 &  1 &  1 &  0 &  0 &  1 &  0 \\ 
24 &  1 &  1 &  0 &  1 &  1 &  1 &  1 &  1 &  1 \\ 
25 &  1 &  1 &  0 &  1 &  1 &  1 &  0 &  1 &  0 \\ 
26 &  0 &  0 &  1 &  1 &  1 &  0 &  1 &  1 &  1 \\ 
27 &  0 &  1 &  1 &  0 &  1 &  0 &  0 &  1 &  0 \\ 
28 &  0 &  1 &  0 &  1 &  1 &  0 &  1 &  1 &  0 \\ 
29 &  0 &  1 &  1 &  0 &  1 &  0 &  1 &  1 &  0 \\ 
30 &  0 &  1 &  0 &  1 &  1 &  0 &  0 &  1 &  0 \\ 
31 &  0 &  1 &  0 &  1 &  0 &  0 &  0 &  1 &  0 \\ 
32 &  0 &  0 &  0 &  0 &  1 &  1 &  1 &  1 &  0 \\ 
33 &  0 &  1 &  0 &  0 &  0 &  0 &  1 &  0 &  0 \\ 
34 &  1 &  0 &  0 &  0 &  1 &  1 &  1 &  1 &  0 \\ 
35 &  1 &  1 &  1 &  1 &  0 &  0 &  1 &  0 &  0 \\ 
36 &  1 &  1 &  1 &  1 &  0 &  1 &  1 &  1 &  1 \\ 
37 &  0 &  1 &  1 &  1 &  1 &  0 &  1 &  0 &  1 \\ 
38 &  0 &  1 &  1 &  1 &  1 &  1 &  1 &  0 &  1 \\ 
39 &  0 &  1 &  0 &  0 &  1 &  1 &  0 &  0 &  1 \\ 
40 &  0 &  0 &  0 &  0 &  1 &  0 &  1 &  0 &  0 \\ 
41 &  0 &  1 &  1 &  0 &  1 &  0 &  1 &  0 &  1 \\ 
42 &  1 &  1 &  1 &  0 &  1 &  0 &  1 &  0 &  1 \\ 
43 &  0 &  1 &  0 &  0 &  1 &  0 &  1 &  0 &  1 \\ 
44 &  1 &  1 &  1 &  0 &  1 &  0 &  1 &  1 &  1 \\ 
45 &  1 &  1 &  1 &  0 &  1 &  0 &  1 &  1 &  1 \\ 
46 &  1 &  0 &  1 &  0 &  1 &  0 &  0 &  0 &  0 \\ 
47 &  0 &  1 &  0 &  0 &  1 &  0 &  0 &  1 &  0 \\ 
48 &  0 &  1 &  0 &  0 &  0 &  0 &  0 &  1 &  0 \\ 
49 &  0 &  1 &  1 &  0 &  1 &  0 &  0 &  1 &  1 \\ 
50 &  0 &  1 &  1 &  0 &  1 &  1 &  0 &  0 &  1 \\ 
51 &  0 &  1 &  1 &  0 &  1 &  0 &  0 &  0 &  1 \\ 
52 &  1 &  1 &  0 &  0 &  0 &  1 &  1 &  0 &  1 \\ 
53 &  1 &  1 &  1 &  1 &  1 &  1 &  1 &  1 &  1 \\ 
54 &  0 &  1 &  1 &  0 &  1 &  0 &  0 &  1 &  1 \\ 
55 &  1 &  0 &  1 &  0 &  0 &  0 &  0 &  0 &  1 \\ 
56 &  1 &  1 &  1 &  0 &  0 &  0 &  0 &  1 &  0 \\ 
57 &  1 &  1 &  0 &  0 &  0 &  1 &  1 &  1 &  1 \\ 
58 &  0 &  1 &  0 &  0 &  0 &  0 &  1 &  0 &  1 \\ 
59 &  0 &  1 &  1 &  0 &  0 &  1 &  1 &  1 &  1 \\ 
60 &  0 &  1 &  1 &  0 &  0 &  1 &  0 &  1 &  0 \\ 
\hline
\end{tabular}
\end{table}
\end{center}

\begin{center}
\begin{tabular}{ |c|c|c|c|c|c|c|c|c|c| } 
\hline
Grid Number & NA=0.1 & NA=0.2 & NA=0.3 & NA=0.4 & NA=0.5 & NA=0.6 & NA=0.7 & NA=0.8 & NA=0.9 \\
\hline
61 &  1 &  1 &  1 &  0 &  0 &  1 &  0 &  1 &  1 \\ 
62 &  1 &  0 &  1 &  0 &  0 &  1 &  0 &  0 &  1 \\ 
63 &  1 &  0 &  1 &  0 &  0 &  1 &  0 &  0 &  1 \\ 
64 &  1 &  0 &  0 &  1 &  0 &  0 &  0 &  0 &  0 \\ 
65 &  0 &  1 &  1 &  1 &  0 &  1 &  1 &  0 &  0 \\ 
66 &  0 &  1 &  1 &  1 &  0 &  0 &  1 &  1 &  0 \\ 
67 &  0 &  1 &  1 &  0 &  1 &  1 &  1 &  1 &  1 \\ 
68 &  0 &  1 &  1 &  0 &  1 &  1 &  1 &  1 &  1 \\ 
69 &  1 &  1 &  1 &  0 &  1 &  0 &  0 &  1 &  0 \\ 
70 &  0 &  0 &  1 &  0 &  0 &  0 &  0 &  1 &  0 \\ 
71 &  1 &  1 &  1 &  1 &  0 &  1 &  0 &  0 &  0 \\ 
72 &  1 &  1 &  1 &  0 &  0 &  0 &  0 &  1 &  0 \\ 
73 &  1 &  1 &  0 &  0 &  0 &  1 &  1 &  1 &  0 \\ 
74 &  0 &  1 &  1 &  0 &  0 &  1 &  0 &  1 &  1 \\ 
75 &  1 &  1 &  1 &  0 &  0 &  0 &  0 &  0 &  1 \\ 
76 &  0 &  1 &  1 &  1 &  0 &  1 &  0 &  0 &  1 \\ 
77 &  1 &  1 &  1 &  1 &  0 &  1 &  0 &  0 &  0 \\ 
78 &  0 &  1 &  1 &  0 &  1 &  1 &  1 &  1 &  0 \\ 
79 &  0 &  1 &  1 &  0 &  1 &  0 &  1 &  1 &  1 \\ 
80 &  1 &  1 &  1 &  0 &  0 &  1 &  0 &  1 &  0 \\ 
81 &  1 &  1 &  1 &  1 &  0 &  0 &  0 &  1 &  0 \\ 
82 &  1 &  0 &  1 &  1 &  0 &  1 &  1 &  0 &  1 \\ 
83 &  1 &  0 &  1 &  1 &  0 &  0 &  1 &  0 &  1 \\ 
84 &  0 &  0 &  1 &  1 &  1 &  0 &  1 &  0 &  1 \\ 
85 &  0 &  0 &  1 &  0 &  0 &  0 &  1 &  0 &  1 \\ 
86 &  1 &  0 &  1 &  1 &  0 &  0 &  1 &  0 &  0 \\ 
87 &  0 &  0 &  1 &  1 &  0 &  0 &  1 &  1 &  1 \\ 
88 &  1 &  0 &  1 &  0 &  0 &  0 &  1 &  0 &  1 \\ 
89 &  1 &  0 &  1 &  1 &  1 &  0 &  1 &  1 &  1 \\ 
90 &  1 &  0 &  1 &  1 &  1 &  0 &  0 &  0 &  1 \\ 
91 &  1 &  0 &  1 &  1 &  1 &  0 &  1 &  0 &  1 \\ 
92 &  0 &  0 &  1 &  1 &  0 &  0 &  1 &  0 &  1 \\ 
93 &  0 &  0 &  1 &  0 &  1 &  0 &  1 &  0 &  1 \\ 
94 &  0 &  0 &  0 &  1 &  1 &  0 &  0 &  1 &  1 \\ 
95 &  1 &  0 &  0 &  1 &  1 &  0 &  0 &  1 &  1 \\ 
96 &  1 &  0 &  0 &  1 &  0 &  1 &  1 &  1 &  0 \\ 
97 &  1 &  0 &  0 &  1 &  1 &  1 &  1 &  1 &  1 \\ 
98 &  0 &  0 &  0 &  0 &  1 &  0 &  1 &  1 &  1 \\ 
99 &  1 &  0 &  0 &  1 &  1 &  0 &  1 &  1 &  1 \\ 
100 &  1 &  0 &  0 &  1 &  1 &  0 &  0 &  1 &  0 \\ 
101 &  1 &  0 &  0 &  1 &  1 &  0 &  0 &  1 &  0 \\ 
102 &  1 &  0 &  1 &  1 &  0 &  0 &  0 &  1 &  0 \\ 
103 &  0 &  0 &  0 &  1 &  1 &  0 &  0 &  1 &  0 \\ 
104 &  1 &  0 &  0 &  1 &  1 &  0 &  0 &  1 &  0 \\ 
105 &  0 &  0 &  0 &  1 &  1 &  0 &  0 &  0 &  0 \\ 
106 &  1 &  0 &  0 &  1 &  1 &  1 &  0 &  1 &  0 \\ 
107 &  1 &  0 &  0 &  1 &  0 &  1 &  0 &  1 &  0 \\ 
108 &  1 &  0 &  1 &  1 &  1 &  1 &  0 &  0 &  0 \\ 
109 &  1 &  0 &  0 &  1 &  1 &  0 &  0 &  1 &  0 \\ 
110 &  0 &  1 &  0 &  1 &  1 &  0 &  0 &  1 &  0 \\ 
111 &  1 &  1 &  0 &  1 &  1 &  1 &  0 &  1 &  0 \\ 
112 &  0 &  0 &  0 &  1 &  1 &  1 &  0 &  1 &  0 \\ 
113 &  1 &  0 &  0 &  1 &  1 &  1 &  1 &  1 &  0 \\ 
114 &  1 &  0 &  0 &  1 &  1 &  0 &  1 &  1 &  0 \\ 
115 &  0 &  0 &  0 &  1 &  1 &  0 &  0 &  1 &  1 \\ 
116 &  1 &  0 &  0 &  1 &  1 &  1 &  0 &  1 &  1 \\ 
117 &  1 &  0 &  0 &  1 &  1 &  1 &  0 &  0 &  1 \\ 
\hline
\end{tabular}
\end{center}

\begin{center}
\begin{tabular}{ |c|c|c|c|c|c|c|c|c|c| } 
\hline
Grid Number & NA=0.1 & NA=0.2 & NA=0.3 & NA=0.4 & NA=0.5 & NA=0.6 & NA=0.7 & NA=0.8 & NA=0.9 \\
\hline
118 &  1 &  0 &  0 &  1 &  1 &  1 &  1 &  0 &  0 \\ 
119 &  0 &  0 &  1 &  1 &  1 &  1 &  0 &  0 &  1 \\ 
120 &  0 &  0 &  1 &  1 &  1 &  0 &  0 &  1 &  1 \\ 
121 &  1 &  0 &  1 &  1 &  1 &  0 &  0 &  1 &  1 \\ 
122 &  1 &  1 &  0 &  1 &  1 &  1 &  0 &  0 &  1 \\ 
123 &  1 &  1 &  0 &  1 &  1 &  0 &  1 &  0 &  1 \\ 
124 &  0 &  0 &  0 &  1 &  1 &  1 &  1 &  0 &  0 \\ 
125 &  1 &  0 &  0 &  1 &  1 &  1 &  0 &  1 &  1 \\ 
126 &  1 &  0 &  0 &  1 &  1 &  1 &  1 &  0 &  1 \\ 
127 &  0 &  0 &  1 &  1 &  1 &  1 &  1 &  0 &  1 \\ 
128 &  1 &  0 &  0 &  1 &  1 &  0 &  0 &  0 &  1 \\ 
129 &  1 &  1 &  1 &  0 &  1 &  1 &  1 &  0 &  1 \\ 
130 &  1 &  0 &  1 &  0 &  1 &  0 &  0 &  0 &  1 \\ 
131 &  1 &  0 &  0 &  0 &  1 &  1 &  1 &  0 &  1 \\ 
132 &  1 &  0 &  0 &  0 &  0 &  1 &  1 &  1 &  1 \\ 
133 &  0 &  1 &  0 &  0 &  0 &  1 &  1 &  1 &  1 \\ 
134 &  0 &  1 &  0 &  0 &  0 &  1 &  1 &  0 &  0 \\ 
135 &  1 &  0 &  0 &  0 &  0 &  1 &  0 &  0 &  1 \\ 
136 &  1 &  0 &  1 &  0 &  0 &  0 &  1 &  0 &  1 \\ 
137 &  1 &  0 &  0 &  0 &  0 &  1 &  1 &  0 &  1 \\ 
138 &  1 &  0 &  1 &  0 &  1 &  1 &  1 &  0 &  1 \\ 
139 &  1 &  0 &  0 &  0 &  1 &  1 &  1 &  0 &  1 \\ 
140 &  0 &  0 &  0 &  0 &  0 &  1 &  0 &  1 &  1 \\ 
141 &  1 &  0 &  1 &  0 &  0 &  1 &  1 &  1 &  0 \\ 
142 &  0 &  1 &  1 &  0 &  0 &  1 &  1 &  1 &  0 \\ 
143 &  1 &  1 &  1 &  0 &  0 &  1 &  1 &  0 &  0 \\ 
144 &  1 &  0 &  0 &  0 &  1 &  1 &  1 &  0 &  0 \\ 
145 &  1 &  1 &  0 &  0 &  1 &  1 &  1 &  1 &  0 \\ 
146 &  0 &  0 &  0 &  0 &  0 &  1 &  1 &  1 &  1 \\ 
147 &  1 &  1 &  0 &  0 &  0 &  1 &  0 &  0 &  0 \\ 
148 &  0 &  1 &  0 &  0 &  0 &  1 &  1 &  1 &  0 \\ 
149 &  1 &  0 &  1 &  0 &  0 &  1 &  1 &  1 &  0 \\ 
150 &  1 &  1 &  1 &  0 &  0 &  1 &  1 &  1 &  1 \\ 
151 &  1 &  0 &  1 &  0 &  0 &  1 &  1 &  0 &  1 \\ 
152 &  1 &  0 &  0 &  0 &  0 &  1 &  1 &  1 &  1 \\ 
153 &  1 &  0 &  1 &  1 &  1 &  0 &  1 &  1 &  0 \\ 
154 &  0 &  0 &  0 &  1 &  1 &  0 &  1 &  1 &  0 \\ 
155 &  1 &  1 &  0 &  1 &  1 &  0 &  1 &  1 &  1 \\ 
156 &  1 &  1 &  1 &  0 &  0 &  0 &  1 &  1 &  1 \\ 
157 &  1 &  0 &  1 &  0 &  0 &  0 &  1 &  0 &  0 \\ 
158 &  1 &  1 &  1 &  0 &  0 &  0 &  1 &  1 &  0 \\ 
159 &  0 &  0 &  0 &  0 &  0 &  0 &  0 &  1 &  1 \\ 
160 &  1 &  0 &  0 &  0 &  0 &  0 &  0 &  1 &  1 \\ 
161 &  1 &  0 &  1 &  0 &  1 &  0 &  0 &  1 &  0 \\ 
162 &  1 &  0 &  1 &  0 &  1 &  0 &  0 &  1 &  1 \\ 
163 &  0 &  0 &  1 &  0 &  1 &  0 &  0 &  1 &  1 \\ 
164 &  1 &  1 &  1 &  1 &  0 &  0 &  0 &  1 &  1 \\ 
165 &  1 &  1 &  1 &  0 &  1 &  1 &  0 &  1 &  1 \\ 
166 &  1 &  0 &  0 &  0 &  1 &  1 &  0 &  1 &  0 \\ 
167 &  1 &  0 &  0 &  1 &  1 &  0 &  0 &  1 &  1 \\ 
168 &  1 &  1 &  1 &  0 &  0 &  0 &  0 &  1 &  1 \\ 
169 &  0 &  0 &  1 &  1 &  0 &  0 &  0 &  1 &  1 \\ 
170 &  1 &  1 &  1 &  0 &  1 &  0 &  1 &  1 &  1 \\ 

\hline
\end{tabular}
\end{center}

\begin{center}
\begin{tabular}{ |c|c|c|c|c|c|c|c|c|c| } 
\hline
Grid Number & NA=0.1 & NA=0.2 & NA=0.3 & NA=0.4 & NA=0.5 & NA=0.6 & NA=0.7 & NA=0.8 & NA=0.9 \\
\hline
171 &  1 &  0 &  0 &  1 &  1 &  0 &  0 &  1 &  1 \\ 
172 &  1 &  0 &  1 &  0 &  1 &  1 &  0 &  1 &  1 \\ 
173 &  0 &  0 &  1 &  0 &  1 &  0 &  0 &  1 &  1 \\ 
174 &  1 &  0 &  0 &  0 &  0 &  0 &  0 &  1 &  1 \\ 
175 &  1 &  1 &  0 &  0 &  1 &  0 &  0 &  1 &  1 \\ 
176 &  1 &  1 &  1 &  1 &  0 &  0 &  0 &  0 &  1 \\ 
177 &  0 &  1 &  1 &  1 &  0 &  0 &  1 &  1 &  1 \\ 
178 &  1 &  0 &  1 &  1 &  0 &  0 &  1 &  0 &  1 \\ 
179 &  1 &  0 &  1 &  1 &  1 &  1 &  0 &  0 &  1 \\ 
180 &  1 &  0 &  1 &  0 &  1 &  1 &  0 &  0 &  1 \\ 
181 &  0 &  0 &  0 &  0 &  1 &  1 &  0 &  0 &  1 \\ 
182 &  1 &  0 &  1 &  0 &  1 &  0 &  0 &  0 &  1 \\ 
183 &  1 &  0 &  1 &  1 &  1 &  0 &  0 &  0 &  1 \\ 
184 &  1 &  0 &  0 &  1 &  0 &  0 &  0 &  0 &  0 \\ 
185 &  1 &  0 &  1 &  0 &  0 &  0 &  1 &  1 &  0 \\ 
186 &  1 &  1 &  1 &  1 &  1 &  0 &  1 &  1 &  0 \\ 
187 &  0 &  0 &  1 &  1 &  1 &  1 &  1 &  1 &  0 \\ 
188 &  1 &  0 &  1 &  1 &  1 &  1 &  1 &  0 &  0 \\ 
189 &  1 &  0 &  0 &  0 &  0 &  1 &  0 &  0 &  1 \\ 
190 &  1 &  0 &  0 &  0 &  1 &  0 &  0 &  0 &  1 \\ 
191 &  1 &  0 &  1 &  0 &  1 &  1 &  0 &  0 &  0 \\ 
192 &  1 &  1 &  1 &  1 &  1 &  0 &  0 &  1 &  0 \\ 
193 &  0 &  1 &  1 &  1 &  1 &  0 &  0 &  0 &  1 \\ 
194 &  0 &  1 &  1 &  1 &  1 &  0 &  0 &  1 &  0 \\ 
195 &  1 &  0 &  1 &  1 &  1 &  1 &  1 &  1 &  0 \\ 
196 &  1 &  1 &  0 &  0 &  0 &  0 &  1 &  1 &  0 \\ 
197 &  1 &  1 &  0 &  0 &  1 &  1 &  1 &  0 &  1 \\ 
198 &  1 &  0 &  1 &  0 &  1 &  1 &  1 &  1 &  1 \\ 
199 &  1 &  0 &  1 &  0 &  1 &  1 &  0 &  1 &  1 \\ 
200 &  1 &  0 &  1 &  1 &  1 &  0 &  0 &  0 &  0 \\ 
201 &  0 &  0 &  1 &  1 &  1 &  0 &  0 &  0 &  1 \\ 
202 &  1 &  0 &  1 &  1 &  1 &  1 &  1 &  1 &  0 \\ 
203 &  1 &  1 &  0 &  1 &  0 &  1 &  1 &  1 &  1 \\ 
204 &  1 &  1 &  0 &  1 &  0 &  0 &  1 &  1 &  1 \\ 
205 &  1 &  1 &  1 &  0 &  1 &  1 &  0 &  0 &  1 \\ 
206 &  1 &  1 &  1 &  1 &  1 &  1 &  0 &  0 &  1 \\ 
207 &  1 &  0 &  1 &  1 &  1 &  1 &  0 &  1 &  0 \\ 
208 &  0 &  1 &  1 &  0 &  1 &  0 &  1 &  1 &  1 \\ 
209 &  1 &  0 &  0 &  1 &  1 &  0 &  1 &  1 &  1 \\ 
210 &  1 &  1 &  1 &  1 &  1 &  1 &  1 &  1 &  1 \\ 
211 &  0 &  0 &  1 &  1 &  1 &  1 &  1 &  0 &  1 \\ 
212 &  1 &  1 &  0 &  1 &  0 &  1 &  0 &  1 &  0 \\ 
213 &  1 &  0 &  1 &  0 &  1 &  1 &  0 &  1 &  1 \\ 
214 &  1 &  1 &  1 &  0 &  1 &  0 &  0 &  1 &  1 \\ 
215 &  1 &  1 &  1 &  0 &  1 &  0 &  1 &  1 &  0 \\ 
216 &  0 &  0 &  1 &  1 &  1 &  1 &  1 &  1 &  1 \\ 
217 &  0 &  0 &  0 &  1 &  1 &  1 &  1 &  0 &  1 \\ 
218 &  1 &  0 &  1 &  1 &  1 &  1 &  1 &  1 &  1 \\ 
219 &  1 &  0 &  1 &  1 &  1 &  1 &  1 &  1 &  1 \\ 
220 &  1 &  1 &  1 &  1 &  0 &  0 &  0 &  1 &  1 \\ 
221 &  1 &  1 &  1 &  0 &  1 &  1 &  0 &  1 &  1 \\ 
222 &  1 &  1 &  1 &  0 &  1 &  1 &  1 &  0 &  1 \\ 
223 &  1 &  1 &  0 &  1 &  1 &  1 &  1 &  1 &  0 \\ 
224 &  0 &  0 &  1 &  0 &  1 &  1 &  0 &  1 &  1 \\ 
225 &  1 &  0 &  1 &  1 &  1 &  0 &  1 &  1 &  1 \\ 

\hline
\end{tabular}
\end{center}

\begin{center}
\begin{tabular}{ |c|c|c|c|c|c|c|c|c|c| } 
\hline
Grid Number & NA=0.1 & NA=0.2 & NA=0.3 & NA=0.4 & NA=0.5 & NA=0.6 & NA=0.7 & NA=0.8 & NA=0.9 \\
\hline
226 &  1 &  1 &  0 &  1 &  1 &  1 &  1 &  1 &  1 \\ 
227 &  1 &  0 &  1 &  1 &  1 &  1 &  1 &  1 &  1 \\ 
228 &  1 &  0 &  1 &  0 &  1 &  1 &  0 &  0 &  1 \\ 
229 &  1 &  1 &  1 &  1 &  0 &  0 &  1 &  1 &  1 \\ 
230 &  0 &  0 &  1 &  1 &  1 &  1 &  1 &  1 &  1 \\ 
231 &  1 &  1 &  0 &  1 &  1 &  1 &  1 &  1 &  1 \\ 
232 &  1 &  1 &  1 &  1 &  1 &  0 &  0 &  1 &  0 \\ 
233 &  1 &  1 &  0 &  0 &  1 &  1 &  1 &  1 &  1 \\ 
234 &  1 &  0 &  0 &  0 &  1 &  1 &  1 &  1 &  1 \\ 
235 &  1 &  0 &  1 &  1 &  1 &  1 &  0 &  0 &  1 \\ 
236 &  0 &  0 &  1 &  1 &  1 &  1 &  1 &  1 &  1 \\ 
237 &  1 &  1 &  1 &  1 &  1 &  0 &  1 &  1 &  1 \\ 
238 &  1 &  0 &  1 &  1 &  1 &  0 &  1 &  1 &  1 \\ 
239 &  1 &  0 &  1 &  0 &  0 &  1 &  1 &  1 &  1 \\ 
240 &  1 &  1 &  0 &  1 &  1 &  1 &  0 &  0 &  1 \\ 
241 &  1 &  1 &  1 &  1 &  1 &  1 &  0 &  1 &  1 \\ 
242 &  0 &  1 &  1 &  0 &  1 &  1 &  1 &  1 &  1 \\ 
243 &  1 &  0 &  1 &  1 &  1 &  0 &  1 &  1 &  1 \\ 
244 &  1 &  0 &  1 &  1 &  1 &  1 &  1 &  1 &  1 \\ 
245 &  1 &  0 &  1 &  1 &  1 &  1 &  1 &  0 &  1 \\ 
246 &  1 &  1 &  1 &  0 &  1 &  1 &  1 &  1 &  1 \\ 
247 &  1 &  1 &  1 &  1 &  0 &  1 &  0 &  1 &  1 \\ 
248 &  0 &  1 &  0 &  1 &  1 &  1 &  1 &  0 &  1 \\ 
249 &  1 &  0 &  1 &  1 &  1 &  0 &  1 &  1 &  1 \\ 
250 &  1 &  1 &  1 &  1 &  1 &  1 &  1 &  1 &  1 \\ 
\hline
\end{tabular}
\end{center}

%